\documentclass[
  reprint,
  amsmath,amssymb,
  aps,
  prd,
  nofootinbib,
]{revtex4-2}

\usepackage{graphicx}
\usepackage{dcolumn}
\usepackage{bm}
\usepackage{framed}
\usepackage{hyperref}
\usepackage{csquotes}

\usepackage{xcolor}
\definecolor{secTwoText}{RGB}{0,105,100}
\definecolor{secTwoNote}{RGB}{145,75,0}

\begin{document}

\title{Lagrangian Bias as a Gaussian Random Field}

\author{Arka Banerjee}
\affiliation{Department of Physics, Indian Institute of Science Education and Research, Pune 411008, India}

\begin{abstract}
Halo or tracer bias is central to understanding cosmological clustering and inferring cosmological parameters.
Halo bias is typically treated as a set of coefficients in a perturbative expansion. Here, we reformulate Lagrangian bias as a field, by establishing that every point in a Gaussian density field has a well-defined scale-independent Lagrangian bias. This field exists before any tracer population is specified, with collapse or selection determining which Lagrangian patches contribute to a population. Halo bias emerges from the appropriate averages over the selected locations, with the predictions tested against $b(M)$ measurements from simulations. Extending the formalism to include additional jointly Gaussian fields in patch selection produces an approximately Gaussian distribution of halo bias at fixed mass, scale-dependent Lagrangian bias, and \enquote{assembly bias}. A concrete toy example demonstrates how a specific selection rule interacts with the underlying bias field to produce bias distributions, whose moments can be connected to the traditional bias coefficients. The construction also yields positive tracer densities everywhere, in contrast to current perturbative bias expansions.
These results provide a foundation for developing realistic statistical collapse models that connect local Lagrangian selections directly to tracer clustering.
\end{abstract}

\maketitle

\section{Introduction}
\label{sec:introduction}
One of the central problems in understanding the Large-Scale Structure (LSS) of the Universe is developing a framework that connects the clustering of observed tracers or special points to the underlying clustering of matter that is governed by various cosmological parameters of interest. Traditionally, this connection has been expressed in terms of tracer bias, usually written as a perturbative expansion of the tracer overdensity in terms of operators defined from the matter overdensity. Eulerian bias approaches perform this expansion using the evolved matter fields and operators, while Lagrangian approaches perform the expansion in terms of the initial fluctuation fields, and then use a prescription to connect to tracer overdensities at late times. These approaches have been formalized at various orders within Lagrangian perturbation theory~\cite[e.g.][]{fry1993,mcdonald2006,mcdonald2009,matsubara2008,matsubara2008b} and the Effective Field Theory of Large Scale Structure (EFTofLSS)~\cite[e.g.][]{chan2012,baldauf2012,sheth2013,senatore2014,mirbabayi2015,desjacques2018}. The coefficients of these expansions depend on the tracer population under consideration. Various approaches exist in the literature that connect these coefficients to the tracer selection, properties, and abundance. This includes Peak-Background Split (PBS) approaches~\cite[e.g.][]{kaiser1984,cole1989,mowhite1996,sheth1999,dekel1999}, properties of peaks and excursion-set models~\cite[e.g.][]{bbks1986,desjacques2010,bond1991,paranjape2013}, and more recently, separate-universe calculations~\cite[e.g.][]{baldauf2011,li2014,wagner2015,barreira2017}. These have been calibrated and tested rigorously against measurements from simulations~\cite[e.g.][]{tinker2010,lazeyras2016}. The bias formalism has been extended to model the tracer densities directly at field level~\cite[e.g.][]{schmittfull2019,wu2022}, rather than for specific statistics. Recently, hybrid approaches, using Lagrangian bias expansion coupled to non-perturbative displacements from simulations, have been developed that extend the reach of the bias models to clustering at smaller scales~\cite[e.g.][]{schmittfull2019,modi2017,kokron2021,angulo2021,pellejero2022}.

Several aspects of tracer clustering have interesting manifestations in the bias formalism. Halos with the same mass but different formation histories have different large scale bias. This \enquote{assembly bias} and the more general \enquote{secondary bias}, the difference in the large scale bias of sub-populations split on any secondary halo property other than mass, have been studied extensively~\cite[e.g.][]{gao2005,wechsler2006,zentner2007,dalal2008,lazeyras2017}. In cosmologies with local primordial non-Gaussianity (PNG), large-scale tracer bias has a contribution that goes as $k^{-2}$, a distinctive signature that can be used to constrain such models~\cite[e.g.][]{dalal2008png,matarrese2008,slosar2008,desjacques2010png}. Beyond population-averaged coefficients, it has been shown that individual halos can be assigned bias estimates using a suitable estimator, and that these estimates vary at fixed mass with location, and secondary properties~\cite[e.g.][]{paranjape2018}. Recent Lagrangian approaches have found approximately Gaussian distributions of the Lagrangian bias for halo populations and related their moments and cumulants to the conventional bias coefficients~\cite[e.g.][]{stucker2025a,stucker2025b}. All these aspects point to the rich phenomenological structure of the bias formalism.

In this paper, we develop a fundamentally different approach to Lagrangian bias. We show that the statistical properties of Gaussian Random Fields (GRFs) allow us to assign a scale-independent local clustering response to every point in the initial conditions. This promotes Lagrangian \enquote{bias} to a field, one that is crucially defined before any tracer population is specified. A collapse or selection prescription selects patches in this field, which, in turn, determines the bias distribution and clustering of the resulting population. Additional operators of the overdensity fields act as further conditioning variables in the language of GRFs and can be used to connect the clustering properties of the selected populations to the fields participating actively in the selection.

We test the new formalism by comparing predictions from the patch-based calculations with the measured large-scale clustering of halo populations. We then quantify how the bias field formalism predicts differences in bias coefficients and the selected bias distribution when selection rules are  based only on local density vs. when they are based on the density and additional operators. Finally, we show an illustrative worked example of how selection rules interact with the bias field to determine the conventional bias coefficients and a tracer field overdensity. The construction yields a tracer field density that is non-negative everywhere, and the positivity is retained after advection to Eulerian space. Together, these results provide a basis for developing statistical collapse prescriptions that predict tracer clustering through their selection of the initial fields.

The paper is organized as follows. In Sec.~\ref{sec:bias_field}, we define the Lagrangian bias field, and develop the required machinery. In Sec.~\ref{sec:tracer_populations}, we relate pointwise and patch estimates to individual halos and also to the population clustering. In Sec.~\ref{sec:halo_assembly}, we examine the role of active selection operators and their connection to assembly bias. In Sec.~V, we work through an explicit toy selection model, and discuss various features. We discuss the implications and future directions in Sec.~\ref{sec:discussion}.

\section{The Lagrangian bias field}
\label{sec:bias_field}

\subsection{Scale-independent pointwise bias}
\label{sec:pointwise_bias}

The overdensity at a point $\mathbf q_0$ in a Gaussian random field of
finite volume $V$ can be written as
\begin{equation}
\delta(\mathbf q_0)=V^{-1}\sum_{\mathbf k}
\tilde\delta(\mathbf k)e^{i\mathbf k\cdot\mathbf q_0}.
\end{equation}
There is an implicit upper limit on the magnitude of the wavevectors appearing in the sum, $k_{\rm max}$, related to the resolution of the field. The Fourier modes have covariance
$\langle\tilde\delta(\mathbf k)\tilde\delta(\mathbf k')\rangle
=V\delta^K_{\mathbf k,-\mathbf k'}P(k)$.
The density $\delta(\mathbf q_0)=\sum_k\delta_k(\mathbf q_0)$ is therefore a
sum of independent Gaussian contributions from disjoint Fourier shells,
\begin{equation}
\delta_k(\mathbf q_0)=V^{-1}\sum_{|\mathbf k|\approx k}
\tilde\delta(\mathbf k)e^{i\mathbf k\cdot\mathbf q_0}.
\end{equation}
Its variance is $\sigma^2=\sum_k\sigma_k^2$, where
$\sigma_k^2=(N_k/V)P(k)$ is the variance of $\delta_k$.
$N_k$ represents the number of modes in the Fourier shell.

\begin{figure*}[t]
  \includegraphics[width=0.9\textwidth]{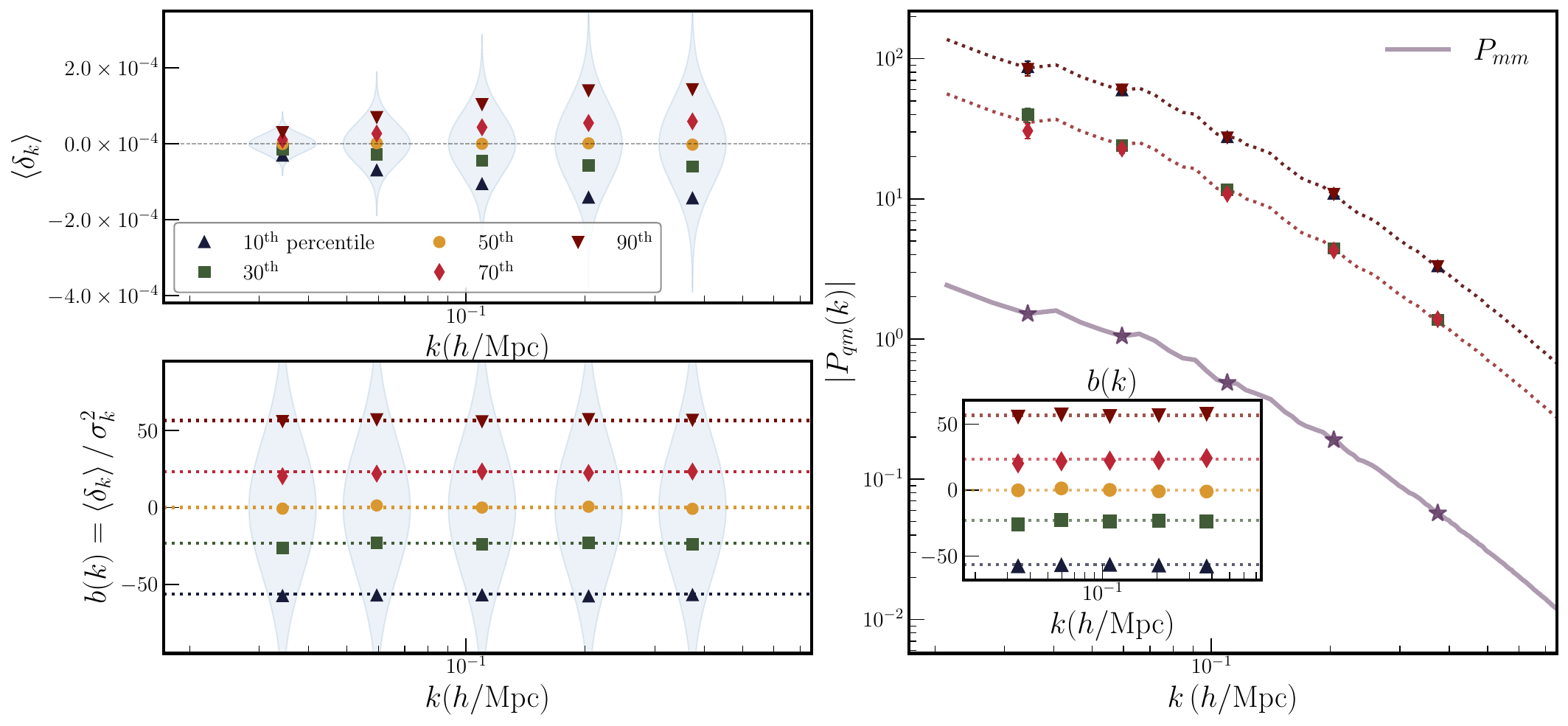}
  \caption{Demonstration of scale-independent pointwise bias using the initial conditions of a \textsc{Quijote} simulation at $z=127$, binned by grid-scale density percentile. \textit{Top left}: The shell-averaged density $\langle\delta_k\rangle_{\mathbf q}$ at $5$ representative $k$-shells varies with $k$. Violins show the width of the underlying distribution. \textit{Bottom left}: After normalizing by $\sigma_k^2$, the conditional mean $\langle\delta_k\rangle_{\mathbf q}/\sigma_k^2$ is the same across all five $k$-shells (symbols), collapsing to the theoretical prediction $\langle \delta(\mathbf{q})/\sigma^2\rangle$ (dotted lines). \textit{Right}: Cross-power spectrum of each density bin with the full matter field. The inset shows the bias $P_{qm}/P_{mm}$; the measured values (symbols) agree with $\langle\delta(\mathbf{q})/\sigma^2\rangle$ (dotted lines) --- exactly the same prediction as in the bottom left panel, illustrating the agreement between the cross-power measurement and the conditional-mean prediction.}
  \label{fig:schematic}
\end{figure*}

Since both $\delta(\mathbf q_0)$ and $\delta_k(\mathbf q_0)$ are Gaussian random variables, the property of multivariate Gaussians gives the conditional mean of $\delta_k(\mathbf q_0)$ at fixed $\delta(\mathbf q_0)$ to be
\begin{equation}
\label{eq:bias_field}
\langle\delta_k\rangle_{\mathbf q_0}
\equiv \langle\delta_k(\mathbf q_0)\mid\delta(\mathbf q_0)\rangle
=\frac{\sigma_k^2}{\sigma^2}\,\delta(\mathbf q_0),
\end{equation}
since by shell independence $\langle\delta_k\delta\rangle=\sigma_k^2$.
Dividing by the shell variance yields 
\begin{equation}
\label{eq:shell_bias}
\frac{\langle\delta_k\rangle_{\mathbf q_0}}{\sigma_k^2}
=\frac{\delta(\mathbf q_0)}{\sigma^2}\equiv b(\mathbf q_0),
\end{equation}
The quantity $b(\mathbf q)$ is therefore well-defined at every point in the field, and predicts a $k$-independent mean contribution from every shell at that point. It follows that $b(\mathbf q)$ itself can be treated as a well-behaved Gaussian random field.

The bias of a set of tracers, usually a collection of points, in cosmology is  defined as the ratio of the cross-power spectrum of the overdensity field defined by the collection of points and the underlying matter density field, $P_{tm}(k)$ and the power spectrum of matter $P_{mm}(k)$, i.e. $b(k) = P_{tm}(k)/P_{mm}(k)$. We now consider the case where we select only one tracer, or a single point $\mathbf q_0$.
Consider the cross-correlation between the overdensity field created by a delta function placed at $\mathbf q_0$, i.e.
$V \delta^{(3)}(\mathbf q-\mathbf q_0) - 1$, and the overdensity field $\delta(\mathbf q)$. For any Fourier shell denoted by $k$, the cross power spectrum is
\begin{equation}
P_{q_0m}(k)=\frac{1}{N_k}\sum_{|\mathbf k|\approx k}
 e^{i\mathbf k\cdot\mathbf q_0}\tilde\delta(\mathbf k)
=\frac{V}{N_k}\,\delta_k(\mathbf q_0).
\end{equation}
Dividing by the power spectrum of the overdensity field $P(k)$ in the same shell yields
\begin{equation}
\hat b_{\mathbf q_0}(k)=P_{q_0m}(k)/P(k)=\delta_k(\mathbf q_0)/\sigma_k^2 \, .
\end{equation}
If the local overdensity at $\mathbf q_0$ is $\delta(\mathbf q_0)$, the conditional mean of this quantity, for any $k$ is
\begin{equation}
\label{eq:point_bias}
\langle \hat b_{\mathbf q_0}(k)\mid\delta(\mathbf q_0)\rangle
=\frac{\langle\delta_k\rangle_{\mathbf q_0}}{\sigma_k^2}=b(\mathbf q_0).
\end{equation}
Conditioning on the local overdensity alone gives the same mean
for the single-tracer cross-power ratio in every shell, and the result equals $b(\mathbf q_0)$ from Eq.~\eqref{eq:shell_bias}. This implies that every point in a realization of a GRF has a well-defined notion of scale-independent \enquote{bias}. As a consequence, Lagrangian bias $b(\mathbf q)$ should be thought of as a field.

Fig.~\ref{fig:schematic} illustrates the equivalence between the GRF result (Eq.~\eqref{eq:shell_bias}) and the standard cosmological definition of bias through power spectra. Using
the initial conditions of a \textsc{Quijote} simulation~\cite{Quijote2020}, we measure the normalized shell means and the
tracer--matter cross-power ratios in bins of grid-scale density percentile
with five representative Fourier shells shown for clarity. The simulation measurements and the analytical estimate using Eqs.~{\eqref{eq:shell_bias},\eqref{eq:point_bias}} all agree with each other.

\subsection{Real-space derivation}
\label{sec:real_space}

The expression for pointwise bias can also be derived in real space. Let $\delta_R(\mathbf q_0)$ be the angularly averaged overdensity on a thin
shell of radius $R$ about $\mathbf q_0$.
For an isotropic field the two-point correlation function $\xi(R)$ sets the covariance of fluctuations of points separated by $R$. Gaussian conditioning gives
\begin{equation}
\langle\delta_R(\mathbf q_0)\mid\delta(\mathbf q_0)\rangle
=\frac{\xi(R)}{\sigma^2}\,\delta(\mathbf q_0).
\end{equation}
For a single tracer placed at $\mathbf q_0$, the real-space
estimator of bias is
$\hat b_R(\mathbf q_0)=\delta_R(\mathbf q_0)/\xi(R)$.
The conditional mean at fixed $\delta(\mathbf q_0)$ is
\begin{equation}
\langle\hat b_R(\mathbf q_0)\mid\delta(\mathbf q_0)\rangle
=\frac{\langle\delta_R(\mathbf q_0)\mid\delta(\mathbf q_0)\rangle}{\xi(R)}
=\frac{\delta(\mathbf q_0)}{\sigma^2}=b(\mathbf q_0).
\end{equation}
Once again, the result is independent of the radius $R$ of the shell, i.e. scale-independent~\cite{frusciante2012}.

\begin{figure*}[t]
  \includegraphics[width=0.95\textwidth]{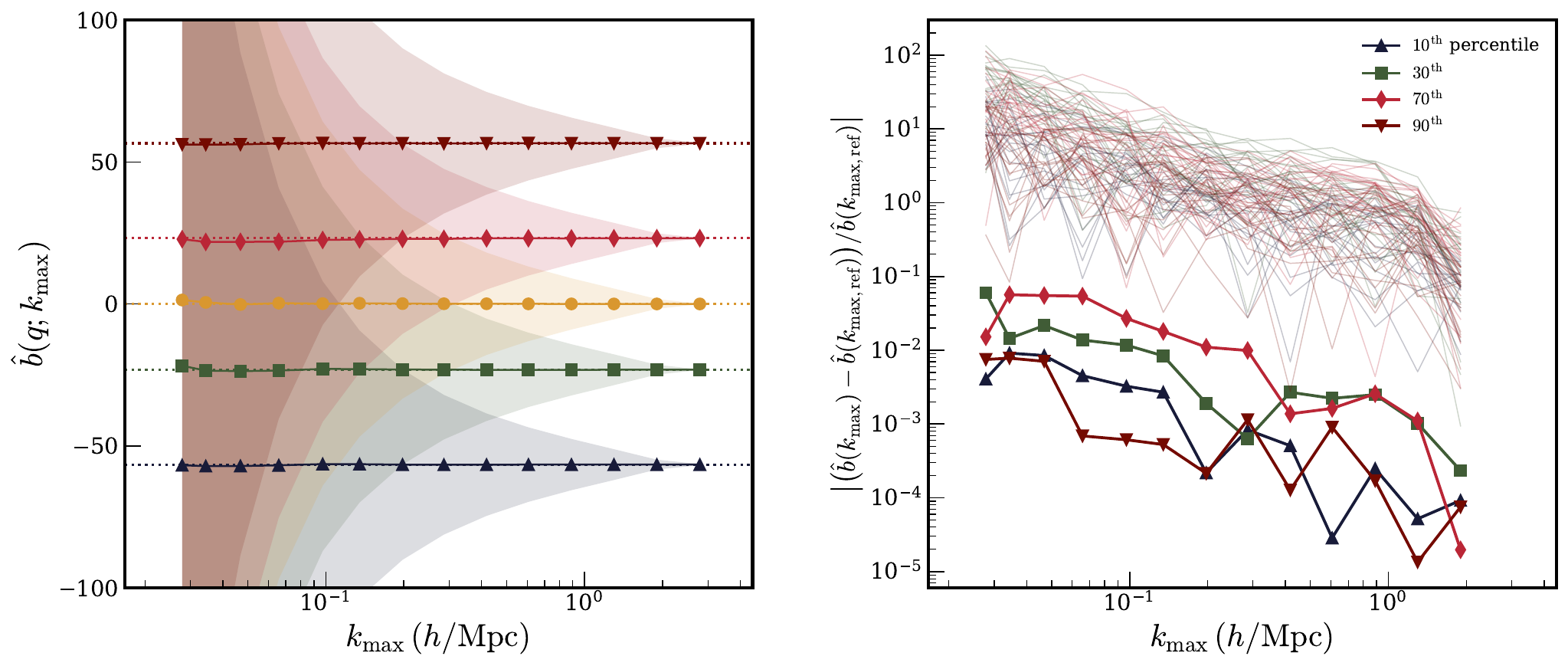}
  \caption{Variation of the bias estimate as a function of $k_{\rm max}$ used in the estimator defined by Eq. \ref{eq:optimal_estimator}. For each percentile bin from Fig. \ref{fig:schematic} , $10000$ points are chosen randomly. The symbols on the left panel represent the mean of the pointwise bias estimates in each bin as $k_{\rm max}$ is varied: the shaded region gives the standard deviation, or scatter across values, in each bin. The right panel plots the convergence of the bias value with respect to the reference $k_{\rm max}\approx3h/{\rm Mpc}$ --- the highest $k$ that can be accessed from the initial conditions grid. The thicker lines represent the mean, while the lighter shaded lines represent the trajectories for a subset of individual points. We leave out the $50$th percentile bin from the right panel since its $b\sim 0$.}
  \label{fig:uvsafety}
\end{figure*}

\subsection{Bias estimator in a single realization}
\label{sec:bias_measurement}
We now address the question of how bias values should be assigned to points in a single realization. At a fixed resolution, we have already shown that $b(\mathbf q_0) = \delta(\mathbf q_0)/\sigma^2$. At the same time, every shell returns an unbiased estimator of the same: $\hat b_{k}(\mathbf q_0) =  \delta_k(\mathbf q_0)/\sigma_k^2$. Any linear combination of the individual estimators $\hat b_w = \sum_k w_k \hat b_k$ with $\sum w_k=1$ is also an unbiased estimator.
Minimizing the variance of $\hat b_w$ with respect to the individual weights $w_k$ yields the optimal estimator. This procedure gives $w_k = \sigma^2_k /\sum \sigma^2_k = \sigma^2_k/\sigma^2$. Substituting the value of $w_k$ in the original expression yields
\begin{equation}
  \hat b_{\rm opt} = \sum_k \frac{\sigma^2_k}{\sigma^2}\frac{\delta_k(\mathbf q_0)}{\sigma^2_k} = \frac{\sum_k \delta_k(\mathbf q_0)}{\sigma^2} = \frac{\delta(\mathbf q_0)}{\sigma^2}.
\end{equation}
The optimal estimator combining information from individual shells therefore reduces to the same pointwise expression as Eq.~\ref{eq:shell_bias}.

We can apply the same procedure to find the optimal estimator while using a subset of the shells $\mathcal K$. Denoting $S_{\mathcal K} = \sum_{k \in \mathcal K} \sigma^2_k$, we get
\begin{equation}
  \label{eq:optimal_estimator}
  \hat b_{\mathcal K}(\mathbf q_0) = \frac{\sum_{k \in \mathcal K} \delta_k(\mathbf q_0)}{S_{\mathcal K}} =\frac{\sum_{\mathbf k\in\mathcal K}\tilde\delta(\mathbf k)
              e^{i\mathbf k\cdot\mathbf q_0}}
       {\sum_{\mathbf k\in\mathcal K}P(|\mathbf k|)}\, .
\end{equation}
This point-by-point estimator turns out to be a generalization of the halo estimator used in \cite{paranjape2018}.
Note that $\mathbb E [\hat b_{\mathcal K}|\delta(\mathbf q_0)] =
\delta(\mathbf q_0)/\sigma^2=b(\mathbf q_0)$
for any fixed subset $\mathcal K$, but there is a scatter in the estimate:
\begin{equation}
  {\rm Var}(\hat b_{\mathcal K}|\delta) = \frac {1}{S_{\mathcal K}}- \frac {1}{\sigma^2} \, .
\end{equation}
When conditioning on the sharp-$k$-smoothed field $\delta_{R_L}$, with $\mathcal K$ a subset of its retained shells, the same result holds with $\sigma^2$ replaced by $\sigma_{R_L}^2$, so that ${\rm Var}(\hat b_{\mathcal K}\mid\delta_{R_L})=1/S_{\mathcal K}-1/\sigma_{R_L}^2$. This will be relevant when we move to halo bias in the next sections.

Although the compact expression $b=\delta/\sigma^2$ makes bias and density appear algebraically interchangeable, their distinction lies in how different scales contribute.  At a given point, any subset of Fourier shells, normalized by its own variance, has the same conditional expectation $b$. This is in contrast to the density field where the obtained value depends explicitly on the modes retained. 

The scatter decreases as more shells are included in the estimate. Assuming that the optimal estimator is applied using all shells up to some $k_{\rm max}$, we investigate the behavior of the estimator, in terms of its convergence to the mean and the scatter in the estimate, in Fig.~\ref{fig:uvsafety}. For each percentile bin in Fig.~\ref{fig:schematic}, we select $10^4$ points at random and vary $k_{\rm max}$, holding the realization and the location of the points fixed. The mean of the estimates is stable in each bin over the full range. The individual estimates also approach their full-grid values as
$k_{\rm max}$ increases.

\subsection{Secondary bias fields}
\label{sec:secondary_fields}
While the preceding discussion has focused on a bias field derived from the overdensity field $\delta(\mathbf q)$, the arguments hold for any GRF. Since the action of a linear operator acting on a Gaussian field yields another Gaussian field, fields such as $\nabla^2 \delta(\mathbf q)$ and $H_{ij} (\mathbf q) = \partial_i \partial_j \delta (\mathbf q) - 1/3 \delta_{ij} \nabla^2 \delta(\mathbf q)$ would also support similar constructions. These linear operators act mode by mode in Fourier space, so the $k$ shell arguments presented in Sec.~\ref{sec:pointwise_bias} apply identically. The \enquote{bias} field for operator $\mathcal O$ is  given by 
\begin{equation}
  \label{eq:b_O_definition}
  b_{\mathcal O}(\mathbf q_0) = \mathcal O[\delta](\mathbf q_0)/\sigma_{\mathcal O}^2 \, .
\end{equation}
Its interpretation, similarly, is as the mean scale-independent response of the modes \emph{of that operator} to its local value, in units of its variance. The interpretation in terms of power spectra is as follows: the shell estimator $\hat b_{\mathcal O}(k)$ at point $\mathbf q_0$ is the ratio of the cross-spectrum of a single tracer (or delta function) at location $\mathbf q_0$ and the field $\mathcal O(\mathbf q)$ and the auto spectrum of $\mathcal O(\mathbf q)$, i.e. $ \hat b_{\mathcal O}(k) = P_{t \mathcal O}(k)/P_{\mathcal O \mathcal O}(k)$. Note that it is not $P_{t\mathcal O}(k)/P_{\delta \delta}(k)$.

We term these secondary bias fields, and they inherit the multivariate Gaussian structure of the underlying operators, $\mathbf C = \langle \mathcal O_i \mathcal O_j\rangle_V$, where the subscript $V$ refers to the fact that the expectation value is computed over the entire volume. Operators involving derivatives of $\delta(\mathbf q)$ upweight the smaller scale modes. Therefore, the convergence of the estimator as a function of $k_{\rm max}$, which now depends on the cumulative variance of $\mathcal O(\mathbf q)$ up to the choice of $k_{\rm max}$ will be different from that demonstrated in the earlier sections. However, in scenarios where a physical smoothing scale $s$ is present in the problem, such as the Lagrangian radius $R_L$ of halos, all estimators return well-defined values.

It is also useful to evaluate the shell estimator at $\mathbf q_0$ while jointly specifying two mean-zero jointly Gaussian variables, $\delta(\mathbf q)$ and some $\mathcal O(\mathbf q)$ at that point. We can construct a field $\mathcal O_{\perp}(\mathbf q)$ 
\begin{equation}
  \label{eq:perp_field}
  \mathcal O_{\perp}(\mathbf q) = \mathcal O (\mathbf q) - \alpha \delta (\mathbf q), \qquad \alpha = \frac{\langle \delta \mathcal O\rangle}{\langle \delta \delta \rangle } \, ,
\end{equation}
which, by construction, has zero covariance with $\delta(\mathbf q)$. When both variables are specified, Eq.~\ref{eq:point_bias} is modified to read
\begin{eqnarray}
  \langle \hat b_{\mathbf q_0} (k)|  \delta(\mathbf q_0), \mathcal O (\mathbf q_0) \rangle &=& \frac{\langle \delta_k \delta\rangle }{\sigma^2_k} b(\mathbf q_0) + \frac{\langle \delta_k \mathcal O_{\perp}\rangle }{\sigma^2_k}b_{\mathcal O_{\perp}}(\mathbf q_0) \nonumber \\ &=&  b(\mathbf q_0) + \frac{\langle \delta_k \mathcal O_{\perp}\rangle }{\sigma^2_k}b_{\mathcal O_{\perp}}(\mathbf q_0) \, ,
  \label{eq:joint_shell_response}
\end{eqnarray}
where we have used $\langle \delta_k \delta \rangle = \langle \delta_k \delta_k\rangle = \sigma^2_k$ and the zero covariance between the two fields. 
Note that this quantity defined in Eq.~\ref{eq:joint_shell_response} is, in general, $k$-dependent, controlled by $\langle \delta_k \mathcal O_{\perp}\rangle/\sigma^2_k$~\cite{Castorina:2016pqq}.

Initially, the bias fields, $b(\mathbf q)$ and $b_{\mathcal O}(\mathbf q)$, all carry the same information as long as they are related by invertible linear transformations at fixed resolution. However, we will demonstrate below that  secondary bias fields $b_{\mathcal O} (\mathbf q)$ capture non-trivial changes in large scale clustering when the collapse criterion depends on operator $\mathcal O(\mathbf q)$ at fixed values of $\delta_s(\mathbf q)$. We term these ``active'' fields, with redundant fields leaving the clustering fully determined by $b(\mathbf q)$ alone. We will examine this in greater detail in Sec.~\ref{sec:halo_assembly}.


\begin{figure*}[t]
  \includegraphics[width=0.8\textwidth]{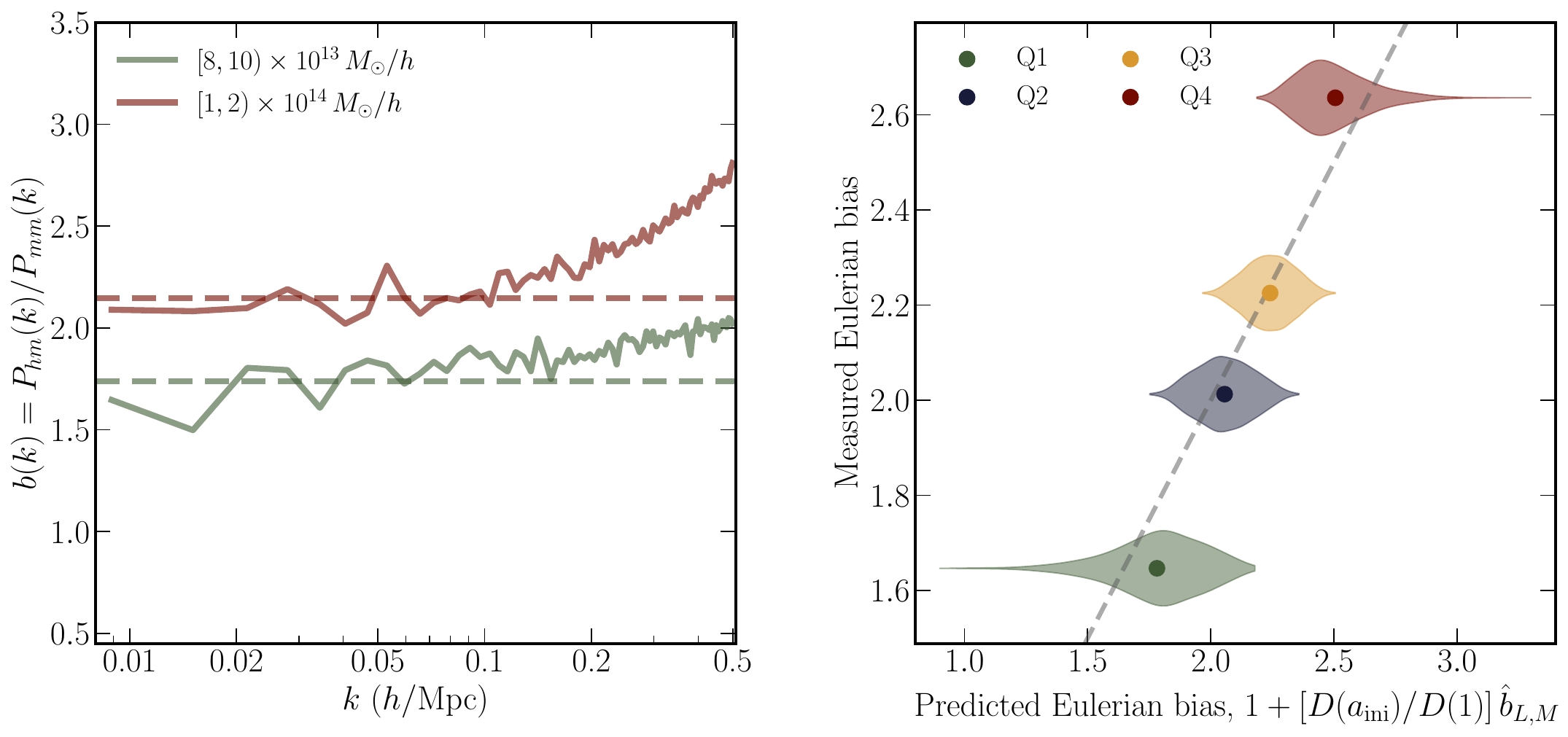}
  \caption{\textit{Left panel}: Measured halo bias $b(k) = P_{hm}(k)/P_{mm}(k)$ for two mass bins at $z=0$ (solid lines). The dashed horizontal lines show the Eulerian bias predicted by Eq.~\eqref{eq:mean_Eulerian_bias}, using the Lagrangian particle estimator of Eq.~\eqref{eq:smoothed_estimator} with $R_L$ set by the median mass in each mass bin. The two agree well on large scales. \textit{Right panel}: Within the higher mass bin, $[1,2] \times 10^{14} M_{\odot}/h$, we further sub-divide into narrow mass bins. Using the predicted $\hat b_{L,M}$, we rank halos within the bin, and divide into 4 quartiles. We pool quartiles across bins, and plot the measured Eulerian bias of each quartile population on large scales (averaged over scales $0.02 \leq k<0.08\,h\,\mathrm{Mpc}^{-1}$) against the prediction for the same collection of objects from Eq.~\eqref{eq:mean_Eulerian_bias}. The dashed line represents the line along which the measured and predicted values match each other. The violin widths represent the distribution of individual halo predictions for comparison. Even within a mass bin, the predicted and measured values track each other. Even in the extreme quartiles, the measurement lies well within the prediction distribution, but the tails make the mean comparison less representative of the distribution.}
  \label{fig:shell_bias}
\end{figure*}

\section{Individual and Population Halo bias from the bias field}
\label{sec:tracer_populations}
Any tracer, or a set of tracers of the matter field can be represented by a collection of representative points. A halo catalog is typically represented by the locations of halo centers --- or by the proto-halo centers in Lagrangian space. In Sec.~\ref{sec:bias_field}, we derived the bias of individual points by examining the ratio of the cross-power spectrum of the tracer field, in this case a single Dirac delta function at the location of the chosen point, and the matter field ($P_{tm}$) and the auto-power spectrum of the matter field ($P_{mm}$). For multiple points, the tracer-matter cross-power spectrum simply averages the contributions of selected points, and therefore the bias of a collection of points is simply the arithmetic mean of the biases of the points, i.e. $\hat b_t (k) = N_t^{-1}\sum_i \hat b_{\mathbf q_i} (k)$. Applying the same variance weighting argument to every tracer, and combining all shells that define the density, we get $b_t = N_t^{-1}\sum_i b(\mathbf q_i)$.

Individual halos, or more correctly, proto-halos, represent coherent and roughly contiguous patches in Lagrangian space with a scale $R_L$ that is typically related to the halo mass. Unlike a collection of points with no intrinsic patch scale, the presence of a physical scale $R_L$ here introduces a subtlety. 
For the former, there is no separation between large and small scales, but in the latter, we can split modes into large-scale modes $k\lesssim 1/R_L$ and short-scale modes $k \gtrsim 1/R_L$. Typically the bias formalism is used to study clustering at scales larger than $R_L$. Applying the optimal estimator from Eq.~\ref{eq:optimal_estimator} for all modes up to $k_{\rm max} \simeq 1/R_L$, it is easy to see that the estimator returns
\begin{equation}
  \label{eq:smoothed_estimator}
  \hat b_{\mathcal K} (\mathbf q_0) = \frac{\sum_{k<1/R_L} \delta_k(\mathbf q_0)}{\sum_{k<1/R_L}\sigma^2_k} = \frac{\delta_{R_L}(\mathbf q_0)}{\sigma^2_{R_L}}\, ,
\end{equation}
where $\delta_{R_L}(\mathbf q)$ is the field smoothed by a sharp $k$-space filter and $\sigma^2_{R_L}$ is the variance of the field.

Applying this estimator over the halo patch, and evaluating the mean yields 
\begin{equation}    
\hat b_{\rm halo}(\mathbf q_c)
= \frac{1}{V_{\rm halo}}\int_{\rm patch} d^3q'\,
  \hat b_{\mathcal K}(\mathbf q')
\approx \frac{1}{N}\sum_{i=1}^{N}\hat b_{\mathcal K}(\mathbf q_i),
\label{eq:individual_halo_bias}
\end{equation}  
where $\hat b_{\mathcal K}(\mathbf q_i)$ are the bias estimates at particle locations in the Lagrangian patch of the halo, and $\mathbf q_c$ represents roughly the proto-halo patch center\footnote{Since we have explicitly filtered out all the sub-patch modes, all quantities of interest will be slowly-varying over the patch. This makes the precise definition of the patch center relatively unimportant in the arguments that follow.}. $V_{\rm halo}$ represents the Lagrangian patch volume. In effect, the patch average applies a window function on the scale of $R_L$, smoothing the bias field. Since $R_L$ is mass-dependent, this smoothing scale is also a function of halo mass. It should be noted that Eq.~\ref{eq:individual_halo_bias} indicates that halos of the same mass, or $R_L$, can have different bias values based on the location of the proto-halo patch, $\mathbf q_c$ in the Lagrangian field. 

On scales that remain linear at
$z=0$, we can also use each halo's Lagrangian patch estimate to predict its
scale-independent Eulerian bias. At linear order, advection from Lagrangian to Eulerian space adds the matter density fluctuation to the tracer overdensity. For a halo whose Lagrangian bias estimate is $\hat b_{\rm halo}(\mathbf q_c)$, the Eulerian bias prediction is simply
\begin{equation}
  \hat b_{E}(\mathbf x_c) = 1 + \frac{D(a_{\rm ini})}{D(1)} \hat b_{\rm halo}(\mathbf q_c) \, ,
\end{equation}
where $D(a)$ is the linear growth factor normalized to $D(1)=1$, and $a_{\rm ini}$ is the epoch
at which the density entering the Lagrangian estimator is normalized
\cite{fry1996}. $\mathbf x_c$ is the Eulerian halo center. The predicted Eulerian bias
varies from halo to halo even at fixed mass. Earlier work has established
halo-to-halo scatter in Eulerian bias estimates~\cite[e.g.][]{paranjape2018}.
We test the mean of our per-halo predictions against the catalogue
cross-power bias below.

We can easily extend the formalism to connect to the mean bias, both Lagrangian and Eulerian for the linear scales, of all halos in a mass range. For $N_h$ halos in a narrow mass range, $M\pm \Delta M$, 
\begin{equation}
  \hat b_{L,M} = \frac{1}{N_h} \sum_{M_h \in M\pm \Delta M} \hat b_{\rm halo}(\mathbf q_{c_h}) \, 
\label{eq:mean_Lagrangian_bias}
\end{equation}
and
\begin{equation}
\hat b_{E,M} = 1 + \frac{D(a_{\rm ini})}{D(1)}\hat b_{L,M} \, .
\label{eq:mean_Eulerian_bias}
\end{equation}
The mean bias in a halo mass bin contains two averages: first over particles in individual halos, and then over each halo in the mass bin (assuming that similar mass halos have roughly the same number of particles associated with them). Predictions for the mean halo bias from this formalism are tested against simulation measurements in the left panel of Fig.~\ref{fig:shell_bias}. On the prediction side, we smoothed the original overdensity field using a slightly tapered $k$-space filter on scale $R_L$ to reduce ringing. Then we assigned every particle a \enquote{bias} value $\delta_{R_L}(\mathbf q)/\sigma_{R_L}^2$ based on their Lagrangian position. From the halo catalog, we identified all particles that were inside the halo boundary at $z=0$, and used Eq.~\ref{eq:mean_Eulerian_bias} to predict the mean bias value for each population. These are plotted using dashed lines in the figure. The solid lines are the Eulerian bias measurements through the ratio of cross and auto power spectra. The dashed lines match the measured large-scale bias well. In the right panel, we further sub-divide the halo population in the higher mass bin into $4$, and show that even within the bin, the predicted Eulerian bias from the current formalism is a good predictor for the measured value. Note that this match to Eulerian space measurements was obtained by using local measurements at scale $R_L$ in Lagrangian space. The shell construction from Sec.~\ref{sec:bias_field} explains why that specific local measurement is able to predict scale-independent large scale bias as measured directly from the simulations. In Appendix~\ref{app:nonhalo_bias}, we demonstrate that the same framework applies not just to halos, but to other collections of patches as well. 

Taken together, these establish a crucial consequence of the bias field framework: that large-scale clustering and collapse are distinct degrees of freedom. The clustering property of all points is already encoded in the initial field of fluctuations before any collapse criterion is applied. On the largest scales the clustering or bias inferred from the initial field predicts the Eulerian bias even at $z=0$ for both halo and non-halo patches. Different collapse criteria pick patches from this existing field, and determine the bias distribution of the selected patches, representing some population of halos or tracers.

\section{Active collapse operators and assembly bias}
\label{sec:halo_assembly}

As pointed out in Sec.~\ref{sec:tracer_populations}, the bias field framework predicts that proto-halos at fixed mass have different $\hat b_{\rm halo}$ depending on the bias field averaged over the proto-halo patch. This defines a distribution $\mathcal P(\hat b_{\rm halo}|M)$, and a natural question is what sets this distribution, or equivalently, the information stored in the distribution. We also showed in Appendix~\ref{app:nonhalo_bias} that one can define the bias of any other patch at the same scale in an analogous manner. Since proto-halos are simply patches in the initial overdensity field that will collapse by $z=0$, $\mathcal P(\hat b_{\rm halo}|M)$ is set by the collapse criterion at mass $M$ and the statistics of the initial field. 

While collapse of individual halos is hard to model, we assume that a collapse model at the population level can be written statistically in terms of Lagrangian GRFs such as $\delta_{R_L}(\mathbf q)$, $\nabla^2 \delta_{R_L}(\mathbf q)$, $s_{R_L}^{ij}(\mathbf q)$, i.e. there exists some complicated function $\mathcal S$
\begin{equation}
  \mathcal S(\mathbf q) = \mathcal S (\delta_{R_L}(\mathbf q), \nabla^2 \delta_{R_L}(\mathbf q), s_{R_L}^{ij}(\mathbf q), ...) \, , 
\end{equation}
such that a particular point $\mathbf q$ lies inside a collapsed patch if $\mathcal S(\mathbf q) = 1$. The underlying fields are jointly Gaussian with known covariance structures. The collapse criterion selects a subset of patches from this field, and the distribution and covariance of the operators in that subset depend on how the selection weights the underlying distribution.

First, let us examine what happens in the case of density-only collapse, focusing on a narrow mass bin, $M_h \in [M - \Delta M, M+ \Delta M]$ with characteristic Lagrangian radius $R_L$. The standard threshold model assigns collapsed mass as the mass enclosed within smoothing scale $s$ for the largest $s$ for which the overdensity at a point crosses a threshold value $\delta_c$. A narrow mass bin selects those regions which cross the threshold value very close to smoothing scale $s=R_L$: regions which cross at a larger smoothing scale are assigned higher masses, and regions crossing at smaller smoothing scales are assigned lower masses. This motivates a single-scale approximation where the proto-halo patch determination depends only on the (sharp $k$-space filtered) overdensity $\delta_{R_L}$ being in a narrow band around $\delta_c$. Assuming that the filtered $\delta_{R_L}$ varies slowly over the patch scale, Eqs.~\ref{eq:smoothed_estimator} and \ref{eq:individual_halo_bias} imply that the bias estimator returns values in a narrow range around $\hat b_{\rm halo} = \langle \delta_{R_L}\rangle_{\rm sel}/ \sigma_{R_L}^2 \simeq \delta_c/\sigma_{R_L}^2$, and this holds for all $k\leq 1/R_L$. Given that the width of the distribution is negligible, the clustering of the entire population, or $\mathcal P(\hat b_{\rm halo}|M)$, therefore, can be summarized by the mean $\langle \hat b_{\rm halo} \rangle = \delta_c/\sigma_{R_L}^2$. Writing $\nu = \delta_c/\sigma_{R_L}$, we get $\langle b \rangle_M \simeq \nu^2/\delta_c$\footnote{Note that the estimator of this quantity has a scatter set by the specific choices of $k$ shells used in the estimate}.
This matches the high $\nu$ limit of the standard Press--Schechter Lagrangian bias $(\nu^2 - 1)/\delta_c$, but with a few differences. The absence of the $-1/\delta_c$ term reflects our single-scale selection approximation, which does not impose the first-crossing condition used to assign masses in the excursion-set calculation. The standard result holds in the limit of $k \ll 1/R_L$, whereas our result holds for all shells $k<1/R_L$ for sharp $k$-space filtering. Finally, and most importantly, the current framework arrives at the answer by using pre-existing bias values, using the patch selection criterion only to determine which parts of the distribution to average over. 

Now consider the more general case where the collapse selection at fixed $R_L$ is allowed to be a function not just of $\delta_{R_L}$ but also of an additional operator $\mathcal O_{R_L}$, i.e. $\mathcal S(\mathbf q) = \mathcal S(\delta_{R_L}(\mathbf q), \mathcal O_{R_L}(\mathbf q))$.  If $\mathcal S$, or collapse probability, changes with $\mathcal O_{R_L}$ at fixed $\delta_{R_L}$, we term $\mathcal O$ as an active collapse operator. Otherwise, $\mathcal O$ is termed as a passive operator. We will now use the idea of secondary bias fields from Sec.~\ref{sec:secondary_fields} to examine the effects of active collapse operators on the clustering of collapsed objects, as captured by $\mathcal P (\hat b_{\rm halo}|M)$. For concreteness, we choose $\mathcal O_{R_L}$ to be $\nabla^2 \delta_{R_L}$ in the subsequent discussion.

In the density-only model considered above, the selected patches lie in a narrow band of constant $\delta_{R_L}$. When additional operators, such as $\nabla^2 \delta_{R_L}$ also enter the collapse criterion, the selected region will in general be tilted in the $\delta_{R_L}-\nabla^2 \delta_{R_L}$ plane, so that the mean selected density varies across curvature bins. These tilts will typically induce a wider projection of these regions in the $\delta_{R_L}-\nabla^2 \delta_{R_L}$ plane onto the $\delta_{R_L}$ axis compared to the density-only case. The broadening along the $\delta_{R_L}$ axis implies a simultaneous broadening of $\mathcal P(\hat b_{\rm halo}|M)$, and $\langle \hat b_{\rm halo} \rangle$ is no longer enough to characterize the clustering of the full population. If the selection band in the $\delta_{R_L}-\nabla^2 \delta_{R_L}$ plane can be approximated by a linear and relatively narrow relation, we expect that the resulting $\delta_{R_L}$ distribution, and therefore, $\mathcal P(\hat b_{\rm halo}|M)$ will be approximately Gaussian. The prediction follows from the joint statistics of the density, curvature, and Lagrangian bias fields, and has been measured directly in $N$-body simulations~\cite{paranjape2018,stucker2025a, stucker2025b}.
\begin{figure*}[t]
  \includegraphics[width=0.85 \textwidth]{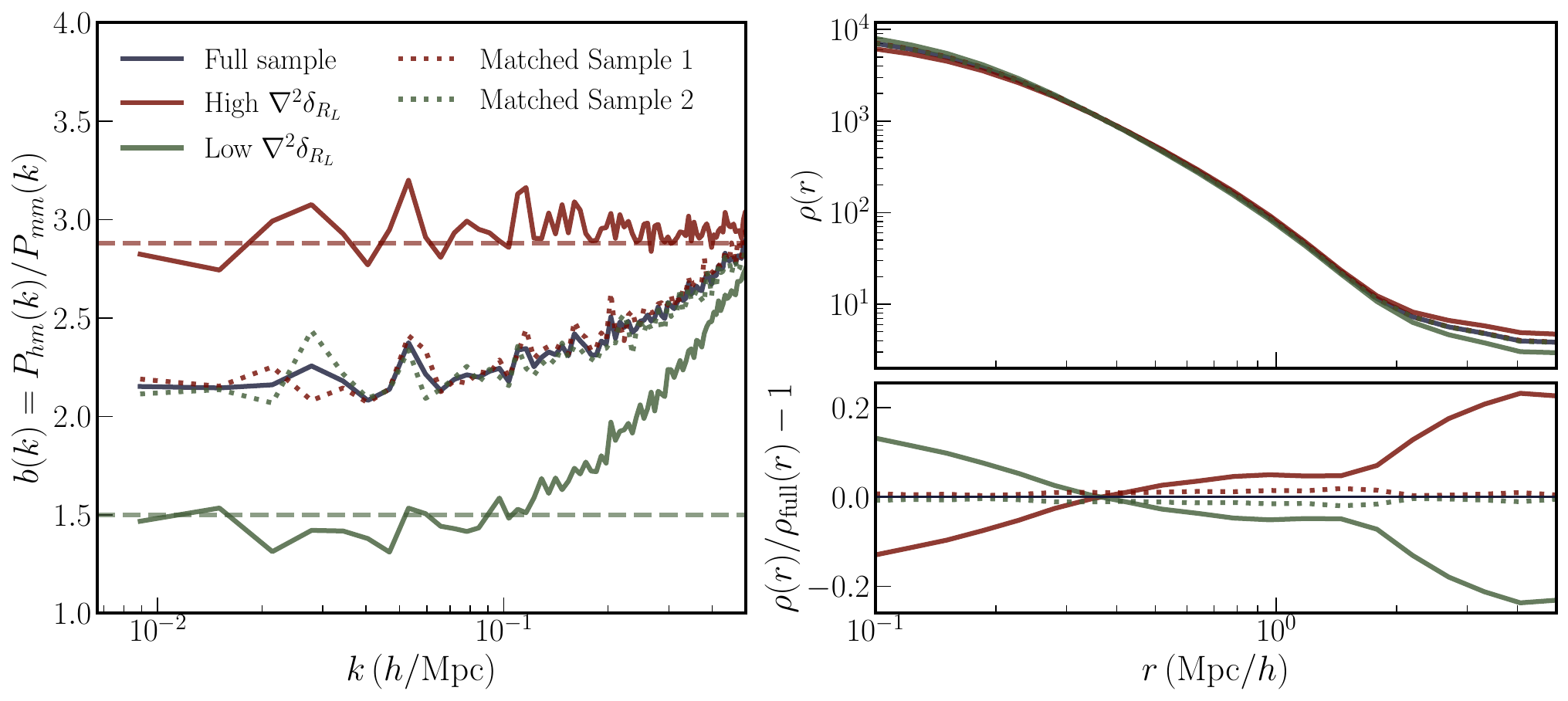}
  \caption{Assembly bias from the secondary bias field $\nabla^2\delta_{R_L}$, for halos in $[1,2] \times 10^{14} M_\odot/h$ at $z=0$. \textit{Left}: Splitting at fixed $\delta_{R_L}$ evaluated over Lagrangian patches of halos by high/low $\nabla^2\delta_{R_L}$ (solid lines) produces a large bias split. The dashed lines represent the mean bias-field prediction of each population evaluated over the halo patch particles using low-$k$ ($k \ll 1/R_L$) modes. Mass-matched samples (dotted lines), which have not been split on $\nabla^2 \delta_{R_L}$ show no bias split. \textit{Right}: The same split produces differences in the stacked density profiles, with the more biased subsample being less concentrated. Both effects are a direct consequence of $\nabla^2\delta_{R_L}$ being an active secondary operator in selecting sub-populations of collapsed objects.}
  \label{fig:assembly_bias}
\end{figure*}

Apart from the broadening of the bias distribution, selection on $\nabla^2 \delta_{R_L}$, or curvature, can also change the scale dependence of the population's Lagrangian bias. Using Eq.~\ref{eq:perp_field}, we can define 
\begin{equation}
  \label{eq:nablasq_perp_def}
  \nabla^2_{\perp} \delta_{R_L}(\mathbf q) = \nabla^2 \delta_{R_L}(\mathbf q) - \alpha \delta_{R_L}(\mathbf q)\, , 
\end{equation}
with 
\begin{equation}
  \alpha =\frac{\langle \delta_{R_L} \nabla^2 \delta_{R_L} \rangle }{ \langle \delta_{R_L}\delta_{R_L} \rangle} \, .
\end{equation}
Eq.~\ref{eq:joint_shell_response} can be written
\begin{equation}
  \label{eq:joint_nablasq}
  \begin{aligned}
  &\bigl\langle \hat b_{\mathbf q_0}(k)
    \mid \delta_{R_L},\nabla^2\delta_{R_L}\bigr\rangle\\
  &\quad = b(\mathbf q_0)
    + \frac{\langle \delta_k\nabla^2_\perp\delta_{R_L}\rangle}
           {\sigma_k^2}\,b_{\nabla^2_\perp}(\mathbf q_0)\\
  &\quad = b(\mathbf q_0)
    + \frac{\langle \delta_k\nabla^2\delta_{R_L}\rangle
          - \alpha\langle \delta_k\delta_{R_L}\rangle}
           {\sigma_k^2}\,b_{\nabla^2_\perp}(\mathbf q_0)\\
  &\quad = b(\mathbf q_0)
    + (-k^2-\alpha)\,b_{\nabla^2_\perp}(\mathbf q_0)\, .
  \end{aligned}
\end{equation}
where 
\begin{equation}
b_{\nabla^2_\perp}(\mathbf q)
\equiv
\frac{\nabla^2_\perp\delta_{R_L}(\mathbf q)}
{\left\langle[\nabla^2_\perp\delta_{R_L}]^2\right\rangle} \, 
\end{equation}
and in the last line of Eq.~\ref{eq:joint_nablasq} we have used the fact that if $\delta_{R_L}(\mathbf q_0)$ receives a contribution of $\delta_k$ from a particular shell, $\nabla^2 \delta_{R_L}(\mathbf q_0)$ receives a contribution $\simeq -k^2 \delta_k$ from the same shell. Averaging this conditional response over locations selected on the values of $\delta_{R_L}$ and $\nabla^2 \delta_{R_L}$ yields
\begin{equation}
  \label{eq:scale_dep_bias}
  b_L(k) = \langle b \rangle_{\rm sel} - (k^2 + \alpha) \langle b_{\nabla^2_\perp} \rangle_{\rm sel} \, ,
\end{equation}
for any shell $k<1/R_L$. If the selection of locations is a function of $\delta_{R_L}$ only, the second term on the RHS is zero since $\nabla_\perp ^2\delta_{R_L}$ is completely independent of $\delta_{R_L}$ by construction and averages to $0$. A collapse criterion which depends actively on the value of $\nabla^2 \delta_{R_L}$ will, in general, have a nonzero mean over $\nabla_\perp ^2\delta_{R_L}$ over the selected locations, and therefore nonzero $\langle b_{\smash{\nabla^2_\perp}} \rangle_{\rm sel}$. This implies that $b_L(k)$ acquires a scale-dependence, and the shape is proportional to $k^2$: the larger the value of $\langle b_{\smash{\nabla^2_\perp}} \rangle_{\rm sel}$, the stronger the effect of this term~\cite{Castorina:2016pqq}. Active collapse operators can, therefore, induce scale-dependent bias in populations, and the shape is set by both the covariances of the underlying operators, and how the active operators select specific locations in the field.

\begin{figure*}[t]
  \includegraphics[width=0.8\textwidth]{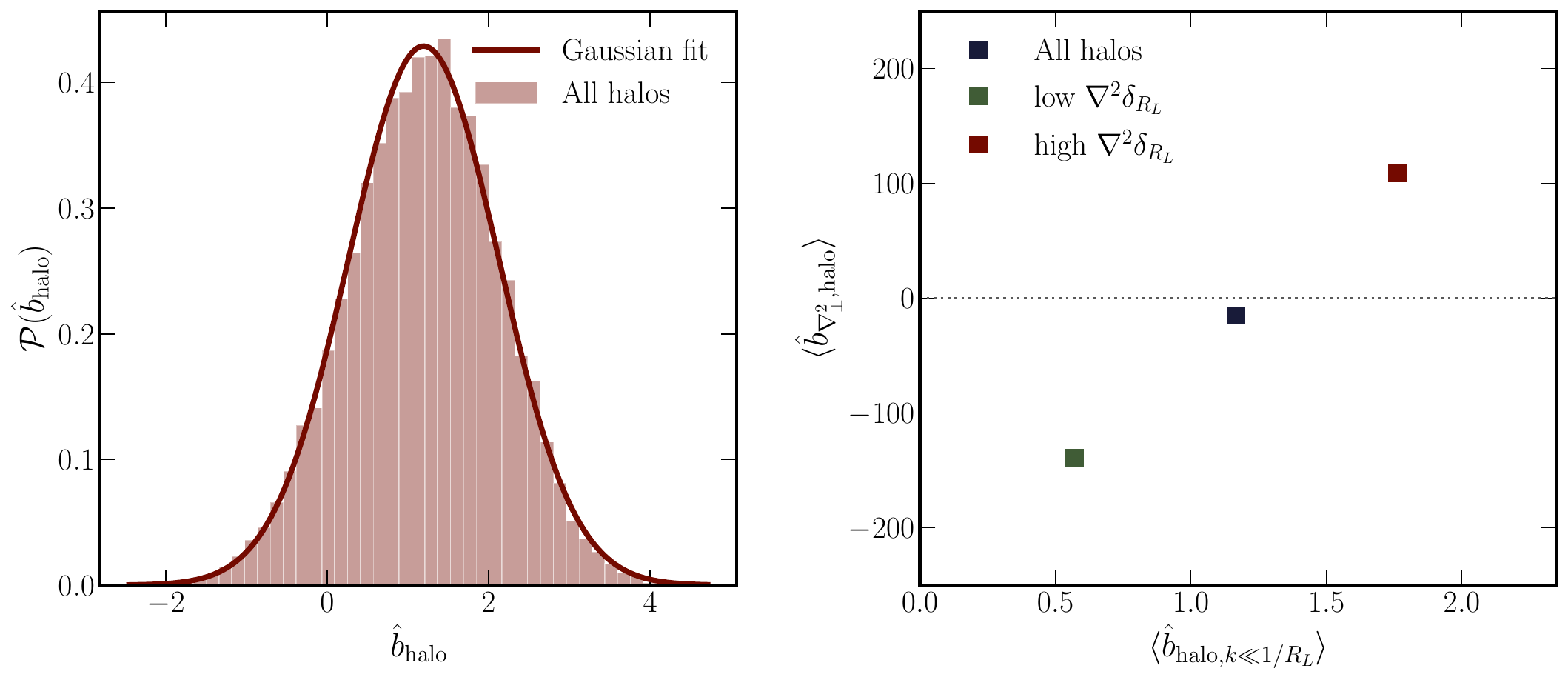}
  \caption{\textit{Left panel:} Distribution of $\hat b_{\rm halo}$ from individual halos in the mass range $[1,2]\times 10^{14} M_{\odot}/h$, estimated using Eqs.~\ref{eq:smoothed_estimator} and \ref{eq:individual_halo_bias}. The histograms represent the measurements while the solid line is the Gaussian fit to the distribution, demonstrating that the distribution is approximately Gaussian. \textit{Right panel:} The $x$-axis plots the mean $\hat b_{\rm halo}$ of a particular population, but only using modes up to $k \sim 0.1 h/{\rm Mpc}$, i.e. large scales. The $y$-axis plots the population mean of $\hat b_{\smash{\nabla^2_\perp}, {\rm halo}}$, using smoothing scale $\sim R_L$. The full population has a small, but nonzero $\langle \hat b_{\smash{\nabla^2_\perp}} \rangle$, which points to a scale-dependent Lagrangian bias, as seen from Eq.~\ref{eq:scale_dep_bias}. Splitting the halos in $\nabla^2 \delta_{R_L}$ (as described in the text), produces two sub-populations with very different mean $\langle \hat b_{\rm halo} \rangle$. The figure, therefore, demonstrates all three signatures of $\nabla^2 \delta_{R_L}$ being an active field in selecting the halo patches in the underlying field. Note that all axes have been multiplied by $D(a_{\rm ini})$ to display Lagrangian bias values at $z=0$.
  }
  \label{fig:opt_bias_scatter}
\end{figure*}

A third measurable manifestation of active collapse operators is the existence of halo sub-populations, selected by the value of $\mathcal O_{R_L}$ at fixed $\delta_{R_L}$, with different large scale bias. To demonstrate this, consider halos in a narrow mass bin, all with Lagrangian sizes given roughly by $R_L$. At fixed $\delta_{R_L}$, we split the halo population into two based on the value of $\nabla^2 \delta_{R_L}$, i.e. a high-curvature sample and a low-curvature sample. Note that this also splits the sample into high and low $\nabla^2_\perp \delta_{R_L}$ samples, since $\alpha$ and $\delta_{R_L}$ are fixed (see Eq.~\ref{eq:nablasq_perp_def}). We can now average Eq.~\ref{eq:joint_nablasq} over the two sub-populations. Since the two sub-populations have the same $\delta_{R_L}$ on average by construction, $\langle b \rangle_{\rm sel}$ will roughly be the same for both. However, since they have very different $\nabla^2_\perp \delta_{R_L}$ distributions, $\langle b_{\smash{\nabla^2_\perp}} \rangle_{\rm sel}$ will be very different. Taking the difference of Eq.~\ref{eq:joint_nablasq} between the two populations, we get 
\begin{equation}
  \Delta b_L(k) \simeq -(k^2 + \alpha) \Delta \Big(\langle b_{\nabla^2_\perp} \rangle_{sel}\Big) \, .
\end{equation}
Even on large scales, $k\ll 1/R_L$ the term proportional to $\alpha$ will persist, and show up as an overall difference in the large scale bias of the two populations. Note that the split can also be present if the mean residual curvature was small for the full population.

We demonstrate this in Fig.~\ref{fig:assembly_bias}, where we consider halos in the mass range $[1,2] \times 10^{14}M_\odot/h$ at $z=0$. By tracing the halo-associated particles back to their Lagrangian positions, we compute the $\langle \delta_{R_L} \rangle$ and $\langle \nabla^2 \delta_{R_L} \rangle$ for each halo by averaging over all particle values in that halo. Additionally, we estimate $\hat b_{\rm halo}$ for each halo by averaging over Eq.~\ref{eq:optimal_estimator} for all particles in that halo, but only using modes with $k\ll 1/R_L$. We divide the halos into bins of narrow $\langle \delta_{R_L}\rangle $, and create two sub-populations by combining halos above (below) the median in $\langle \nabla^2 \delta_{R_L}\rangle$ from each bin. Using this split determined by properties of the initial Lagrangian patch, we measure the $z=0$ bias of the two populations in the left panel of Fig.~\ref{fig:assembly_bias} using the solid lines. There is a clear split in the large-scale bias of the two samples. The dashed lines show the predictions for the Eulerian bias of the two populations from the measurements of $\langle \hat b_{\rm halo}\rangle$ in each population. The bias field predictions and measurements are a good match, but for this, it was crucial to use only $k\ll 1/R_L$ modes in the estimator. The right panel shows that the two sub-populations also differ in their stacked density profiles, with the more strongly clustered population demonstrating a lower average concentration. Together, these trends are consistent with the well-explored phenomenon of high-mass assembly bias~\cite{dalal2008}. In the bias field language, \enquote{assembly bias} is a reflection of additional operators participating actively in the selection of collapsed patches in the underlying bias field. In
Appendix~\ref{app:nonhalo_secondary}, we demonstrate the argument is not
specific to halos, reinforcing the argument that clustering and selection (collapse condition for halos) are decoupled in the initial conditions.

The bias field framework shows that three seemingly distinct large scale clustering phenomena --- the Gaussian width of $b_{\rm halo}$ at fixed $M_{\rm halo}$, a $k^2$ scale-dependence in Lagrangian bias, and assembly bias --- are manifestations of a single central physical idea: the active participation of additional fields, apart from $\delta_{R_L}$, in selecting relevant patches. This selection, along with the statistical properties of the underlying fields, determines the exact nature of all three clustering phenomena. This common origin is difficult to establish from previous, separate frameworks in the literature. Fig.~\ref{fig:opt_bias_scatter} demonstrates all three aspects: the left panel shows the approximately Gaussian distribution of $\hat b_{\rm halo}$ in the mass range $[1,2]\times 10^{14} M_\odot/h$. The right panel shows that the full halo population has a measurable $\langle b_{\smash{\nabla^2_\perp}}\rangle$, which by Eq.~\ref{eq:scale_dep_bias}, implies this population will have a $k^2$ scale-dependent term in its Lagrangian bias. Third, sub-populations split on $\nabla^2\delta_{R_L}$ at fixed $\delta_{R_L}$ over the proto-halo patch produce a clear separation in both $\langle \hat b_{\rm halo} \rangle$ and $\langle b_{\smash{\nabla^2_\perp}}\rangle_{\rm sel}$.

A scientifically interesting scenario in which the \enquote{active} vs. \enquote{passive} field distinction is important is that of local primordial non-Gaussianity. $\phi_{R_L}(\mathbf q)$ is a passive or redundant field for Gaussian initial conditions as it does not appear explicitly in the collapse criterion. Therefore, $\langle b_\phi\rangle$ as well as the full distribution of $b_\phi$ for a set of tracers is set by the correlation of the field $\phi$ with active directions such as $\delta_{R_L}$ and $\nabla^2 \delta_{R_L}$. For cosmologies with local primordial non-Gaussianity, i.e. $f_{\rm NL} \neq 0$, $\phi(\mathbf q)$ plays two different roles: it introduces weak non-Gaussianity in the initial overdensity field $\delta(\mathbf q)$, and it actively participates in setting the collapse criterion~\cite{dalal2008png}. From Eqs.~\ref{eq:joint_shell_response} and \ref{eq:joint_nablasq}, replacing $\nabla^2 \delta_{R_L}$ by $\phi_{R_L}$, it is easy to see that within the bias field formalism, this leads to a scale-dependent bias of the form $1/k^2$. This particular scale-dependent bias signature presents one of the most promising ways of constraining such models~\cite{dalal2008png, matarrese2008, slosar2008, desjacques2010png, schmidt2013}. The formalism presented here also predicts that the rotation of the collapse volume in the $\delta_{R_L}-\phi_{R_L}$ plane, under usual collapse scenarios, can lead to a wider distribution of $\mathcal P(\hat b_{\rm halo}|M)$ in these cosmologies, a novel and possibly measurable signature of $f_{\rm NL}\neq0$ . We leave a quantitative exploration of this scenario to future work.

We close this Section by noting that while the presentation here has focused on collapse being governed by $\delta_{R_L}$ and one additional field or operator, realistic collapse or selection models can involve more. While the formalism presented here extends to such scenarios by increasing the set of conditioning variables, the challenge will be to identify the right fields. Different sets of fields may be active and important at different halo mass scales: for example, tidal shear is expected to be important for collapse in lower mass halos, but is expected to play a smaller role in determining collapse rules for halos at the high mass end of the mass function.

\section{Toy Model Demonstration}

In this Section, we demonstrate a working example of how a  selection rule interacts with the statistics of the initial underlying fields, including the Lagrangian bias field introduced in this paper, to determine the clustering of selected regions that is valid on large scales even at $z=0$. We intentionally keep the selection rule extremely simple in the demonstration --- it is not meant to be a realistic one for halos in any mass range. We focus instead on how this rule selects specific regions in the underlying bias field, and how the traditional bias parameters emerge from the statistics of the bias field over the selected regions. We also demonstrate additional attractive features, such as guaranteed tracer field positivity, that arise from the framework, and differentiate it from existing biasing frameworks in the literature.

\begin{figure}[t]
  \includegraphics[width=0.8\columnwidth]{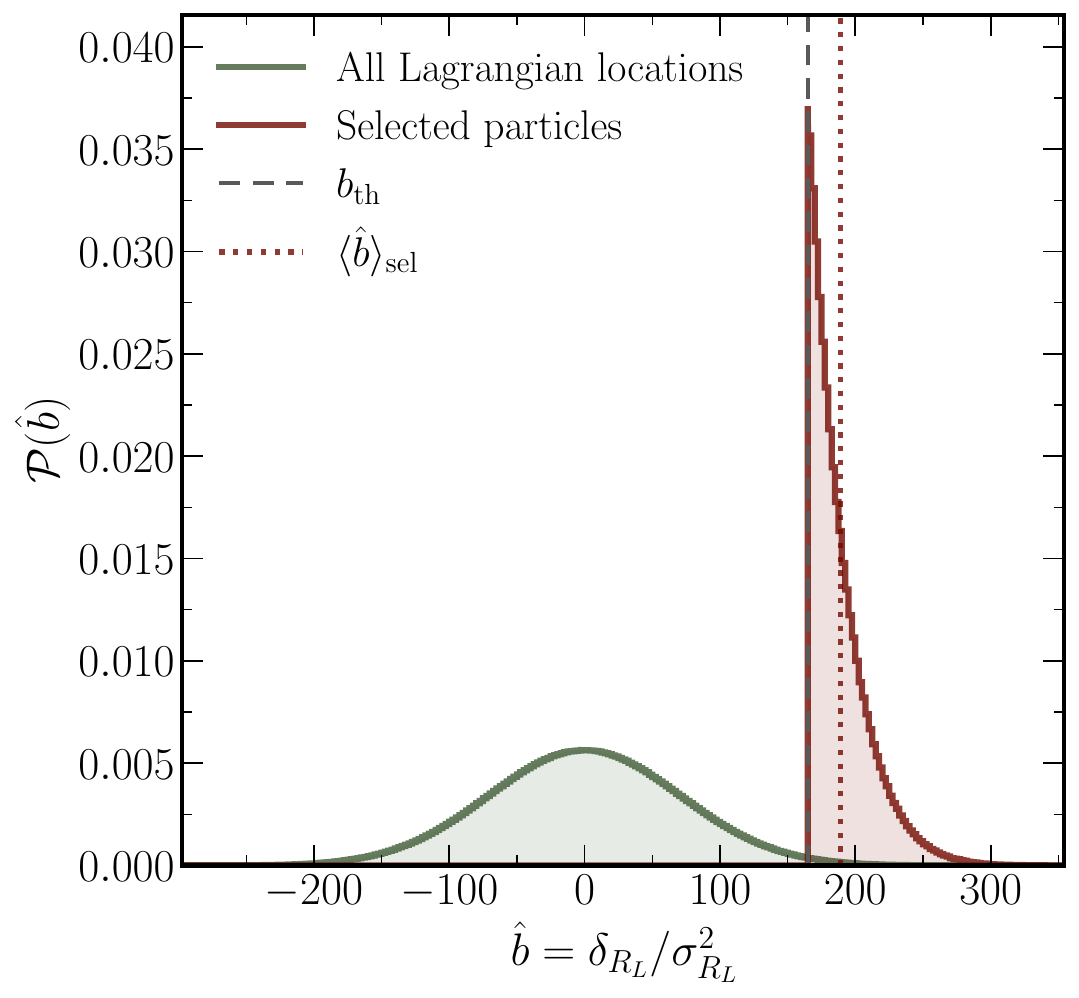}
  \caption{Distribution of the initial density bias field $b_\delta=\delta_{R_L}/\sigma_{R_L}^2$ over all Lagrangian sites (green) and over the upper 1\% selected by the $R=3\,h^{-1}\mathrm{Mpc}$ density threshold (red). Each histogram is independently normalized to unit area. The dashed line represents the bias threshold cut $b_{th} = \delta_{R_L,th}/\sigma^2_{R_L}$. The dotted line represents the mean of the bias distribution of the selected population.}
  \label{fig:threshold_bias_distribution}
\end{figure}

The rule we will use to select particles or regions in the initial density field is a threshold-based one: on a Lagrangian smoothing scale of $R_L = 3 \, {\rm Mpc}/h$\footnote{We use a real-space $3$-dimensional tophat filter for this example.}, we introduce a threshold which selects only the top $1$ percentile of particles based on their Lagrangian $\delta_{R_L}$ value. We denote the resulting overdensity threshold as $\delta_{R_L,th}$. At the Lagrangian position of each simulation particle, we evaluate the bias field $\hat b(\mathbf q)=\delta_{R_L}(\mathbf q)/\sigma_{R_L}^2$, where $\sigma_{R_L}^2=\langle\delta_{R_L}^2\rangle$ is the full-field variance of the same top-hat-smoothed density field. Since we have used the same smoothing scale, we expect a bias selection that is also thresholded: $b_{th} = \delta_{R_L,th}/\sigma^2_{R_L}$. This is shown in Fig.~\ref{fig:threshold_bias_distribution} which shows both the underlying population distribution of $\mathcal P (\hat b)$ in green, as well as the threshold-selected population in red. The selected $\mathcal P(\hat b)$ distribution is far from a Gaussian distribution.

Given the full $\mathcal P_{\rm sel}(\hat b)$ distribution, we can compute various moments of the bias distribution,
\begin{equation}
\mu_n=\int d\hat b\,\hat b^{\,n}\mathcal P_{\rm sel}(\hat b),
\qquad n=1,2,3,\ldots\,.
\label{eq:selected_bias_statistics}
\end{equation}
The moments are then directly connected to the Lagrangian bias coefficients in the following way~\cite{szalay1988,stucker2025b}:
\begin{equation}
\begin{aligned}
b_1^{\rm L}&=\mu_1,\\
b_2^{\rm L}&=\mu_2-\frac{1}{\sigma_{R_L}^2},\\
b_3^{\rm L}&=\mu_3-\frac{3\mu_1}{\sigma_{R_L}^2}.
\end{aligned}
\label{eq:threshold_distribution_bias}
\end{equation}

\begin{table}[t]
\caption{Growth-rescaled Lagrangian bias coefficients from the selected bias
distribution (Eq.~\ref{eq:threshold_distribution_bias}) and the  response calculation (Eq.~\ref{eq:threshold_response_bias}) at the measured threshold.}
\label{tab:threshold_bias_hierarchy}
\begin{ruledtabular}
\begin{tabular}{crr}
$n$ & \shortstack{Bias field\\$D^n(a_{\rm ini})b_n^{\rm L}$}
& \shortstack{Response\\$D^n(a_{\rm ini})b_n^{\rm L}$}\\
\hline
1 & $1.8698$ & $1.8693$\\
2 & $3.0523$ & $3.0505$\\
3 & $4.0640$ & $4.0593$\\
4 & $3.6352$ & $3.6255$\\
5 & $-0.0544$ & $-0.0691$\\
6 & $-7.2276$ & $-7.2404$\\
\end{tabular}
\end{ruledtabular}
\end{table}

\begin{figure*}[t]
  \includegraphics[width=0.8\textwidth]{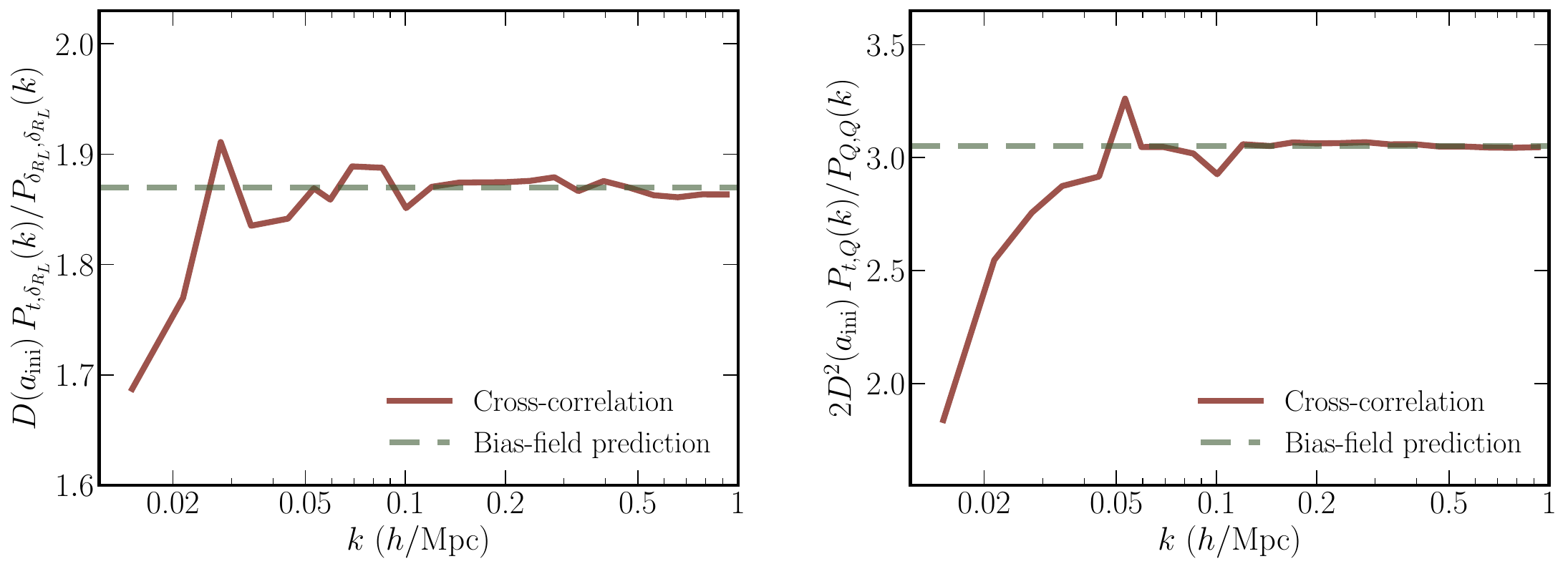}
  \caption{Comparison between the predicted and measured values of $b_1^L$ and $b_2^L$ for the tracer population selected by the density threshold criterion at $R_L = 3\, {\rm Mpc}/h$. The left panel shows the measurement for $b_1^L$ using power spectra measurements on the selected tracer field and $\delta_{R_L}(\mathbf q)$ with the solid line while the prediction is shown with the dashed line. The right panel shows the same for $b_2^L$ computed from the tracer field and $Q(\mathbf q) = \delta^2_{R_L}(\mathbf q) -\langle \delta_{R_L}^2\rangle$. Both measurements and predictions have been multiplied by appropriate powers of $D(a_{\rm ini})$ to make the y-axes comparable.
  }
  \label{fig:cross_bias}
\end{figure*}

We can also compute the usual Lagrangian bias parameters from the traditional response calculations~\cite[e.g.][]{schmidt2013,lazeyras2016}, writing
\begin{eqnarray}
  \label{eq:delta_t_def_trad}
  1 + \delta_t = \frac{f(\delta_l)}{f(0)}
=1+\sum_{n\geq1}\frac{b_n^{\rm L}}{n!}\delta_l^n, 
\end{eqnarray}
where $\delta_t$ is the large-scale tracer overdensity in Lagrangian space, $\delta_l$ represents the long-wavelength mode, and $f(\delta_l)$ is the tracer density as a function of $\delta_l$. Note that the expression is organized as a perturbative expansion around $\delta_l= 0$. Eq.~\ref{eq:delta_t_def_trad} can be inverted to define 
\begin{equation}
b_n^{\rm L}=\frac{1}{f(0)}
\left.\frac{d^n f(\delta_l)}{d\delta_l^n}\right|_{\delta_l=0}.
\label{eq:threshold_response_bias}
\end{equation}
For the simple threshold model here, holding $\delta_{R_L,th}$ fixed, 
\begin{equation}
  f(\delta_l) = \frac 1 2 \mathrm{erfc}\Bigg (\frac{\delta_{R_L,th} - \delta_l}{\sqrt 2 \sigma_{R_L}} \Bigg) \, ,
\end{equation}
the bias coefficients can be evaluated
analytically. Table~\ref{tab:threshold_bias_hierarchy} compares these response
coefficients with those obtained from the selected bias distribution up to
sixth order\footnote{We use the response convention to compare values: the conventions otherwise differ by a factor of $n!$.}. As should be true for a density-only selection~\cite{szalay1988}, the two columns match closely at almost all orders, with a slight difference at $n=5$, which happens to be numerically small. While the two calculations yield similar results, it is worth noting the difference in the two approaches: in the bias field approach, we treat the bias at each Lagrangian location as a pre-existing quantity. The application of the selection rule yields an actual distribution of bias values by selecting specific regions, and the moments\footnote{See \cite{stucker2025b} for arguments why cumulants are a more efficient compression of the Lagrangian clustering.} (or equivalently $b^L_n$) capture properties of this distribution. The response calculation, on the other hand, adds terms organized in a perturbative expansion, and bias coefficients are to be interpreted as the response coefficient at each order, rather than as a summary of an underlying distribution.

Fig.~\ref{fig:cross_bias} demonstrates the agreement between the bias field predictions, and the measurement of $b_1^L$ and $b_2^L$ from the ratio of the cross to the auto power spectrum computed from the tracer field defined by our selection criterion and the operator fields $\delta_{R_L}(\mathbf q)$ and $Q(\mathbf q) = \delta^2_{R_L}(\mathbf q) -\langle \delta_{R_L}^2\rangle$. The predictions match the measurements well down to the small scales in Lagrangian space. Note that in the bias field formalism, the prediction for $b_2^L$ emerged from the first two moments of the distribution in $\mathcal P_{\rm sel}(\hat b)$, i.e. related to a property calculated from the linear $\delta_{R_L}$ field. This is in contrast to the traditional bias expansion where $b_2^L$ is treated as arising from an independent field in the bias expansion, i.e., $\delta^2_{R_L}(\mathbf q) -\langle \delta_{R_L}^2\rangle$.

\subsection{Tracer field positivity}
\label{sec:tracer_field_postivity}

In addition to a distribution of bias values, the selection rules also specify a tracer field. For the simple threshold model, we can define a field $S(\mathbf q) = \Theta(\delta_{R_L}(\mathbf q)- \delta_{R_L,th})$, where $\Theta$ represents the Heaviside theta function, indicating whether a point $\mathbf q$ is selected. The tracer overdensity field is, therefore, defined as
\begin{equation}
1+\delta_t^{\rm L}(\mathbf q)
=\frac{S(\mathbf q)}{\langle S\rangle},
\label{eq:positive_selection_field}
\end{equation}
where the average is over the entire Lagrangian volume, or equivalently, all particle positions in the Lagrangian space. The RHS of Eq.~\ref{eq:positive_selection_field} is clearly $\geq 0$, and therefore $\delta^L_t(\mathbf q) \geq -1$ everywhere. The property holds even when the selection rule is more complicated, and depends on the values of multiple smoothed fields at a given location.

Importantly, the same positivity property holds when we approximate the selection rule by a probabilistic model whose parameters can be inferred. In general, let $p_{\boldsymbol\theta}(\mathbf q)$ be the probability of selecting $\mathbf q$ given the local fields. Assuming $0 \leq p_{\boldsymbol\theta} \leq 1$ over the entire range of parameter choices, the tracer overdensity field is given by 
\begin{equation}
1+\delta_{t,\boldsymbol\theta}^{\rm L}(\mathbf q)
=\frac{p_{\boldsymbol\theta}(\mathbf q)}
{\langle p_{\boldsymbol\theta}\rangle}.
\label{eq:positive_probabilistic_field}
\end{equation}
Once again, the RHS is $\geq 0$ by construction, and therefore $\delta_{t,\boldsymbol\theta}^L$ is bounded below by $-1$~\cite{stucker2025a}. It is also simple to translate the probabilistic selection into changes in the selected bias distribution, since the underlying bias values in the Lagrangian field do not depend on the selection rule. In particular,
\begin{equation}
\mathcal P_{\rm sel}(\hat b\mid\boldsymbol\theta)
=\frac{\left\langle p_{\boldsymbol\theta}(\mathbf q)\,
\delta_{\rm D}[\hat b-\hat b(\mathbf q)]\right\rangle}
{\langle p_{\boldsymbol\theta}\rangle},
\label{eq:probabilistic_selected_bias_distribution}
\end{equation}
where the Dirac delta in the expression simply picks up the bias values at appropriate locations weighted by local probability for selection. The equation above defines the distribution, while the moments can be computed as $\mu_n(\boldsymbol\theta)=
\langle p_{\boldsymbol\theta}\hat b^n\rangle/
\langle p_{\boldsymbol\theta}\rangle$. The distribution and the bias coefficients, through the relevant moments, now explicitly depend on the selection model and its parameters $\boldsymbol\theta$, and these can be varied to approximate the \enquote{true} selection rule. However, the positivity of the tracer field is preserved, and this is crucial when building a realistic forward model of collapse or selection to match the clustering of a particular set of tracers.

\begin{figure*}[t]
\centering
\includegraphics[width=0.98\textwidth]{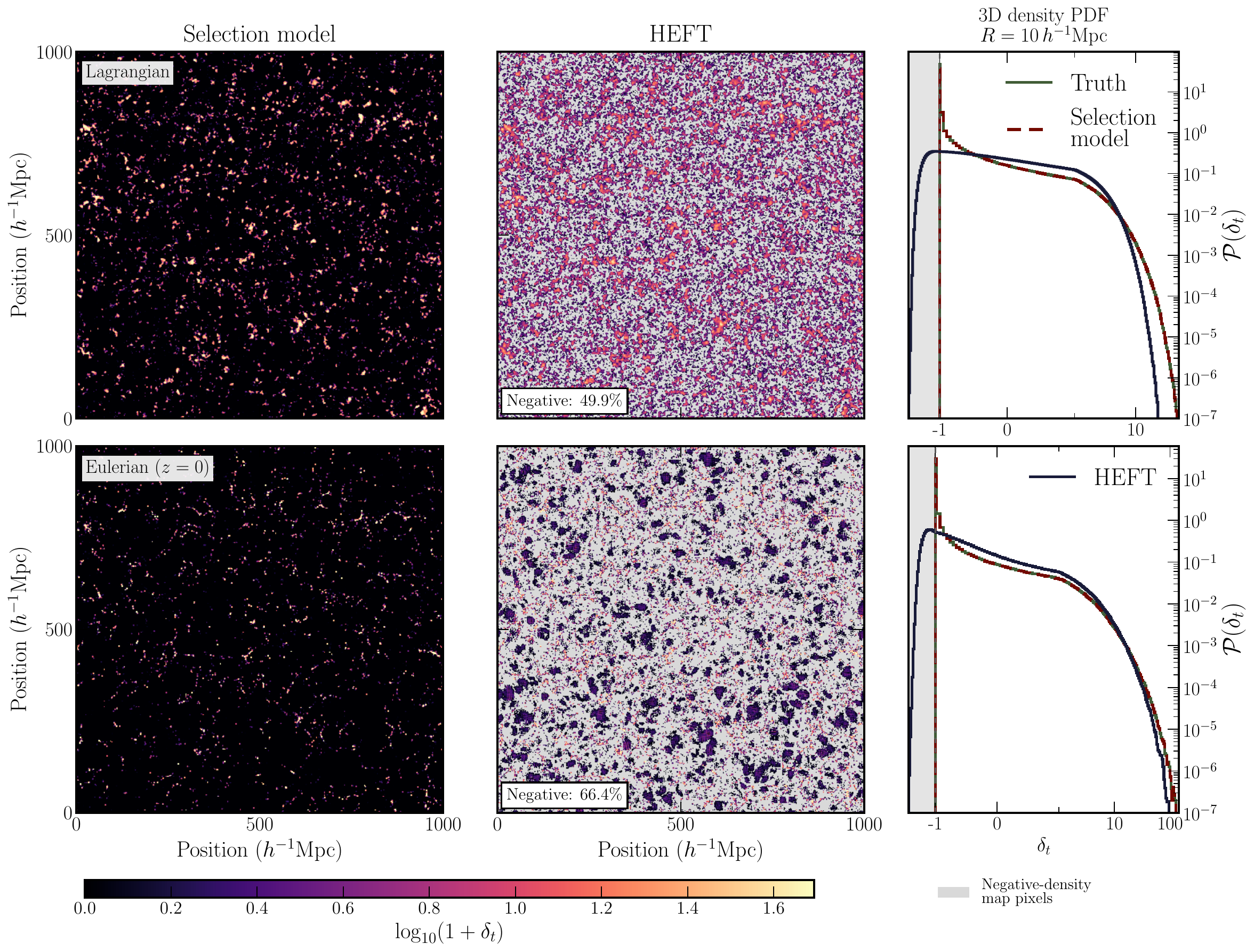}
\caption{\textit{Left panels:} Tracer field positivity at the pixel level for the probabilistic selection model in Eq.~\ref{eq:logistic_recovery_selection} when fit to the true distribution in Eq.~\ref{eq:fuzzy_truth_selection}. The upper panel represents Lagrangian space projected densities while the bottom panel represents Eulerian densities. The projection depth is $15.6\, {\rm Mpc}/h$, with no additional three-dimensional smoothing of the left and middle maps. \textit{Middle panels:} The projected fields no longer have the positivity property under the best-fit HEFT model (Eq.~\ref{eq:heft_lagrangian_tracer_overdensity}). The fraction of negative pixels, shown in gray, in Lagrangian (upper) and Eulerian (lower) panels is indicated.  \textit{Right panel:} Histograms of the $3$-d overdensity distributions, smoothed at $10 \, {\rm Mpc}/h$, in Lagrangian (upper panel) and Eulerian (lower panel) space. Both the true tracer distribution and the probabilistic fit model distribution truncate at $\delta_t = -1$, while the expansion-based Lagrangian bias model has a prominent tail below that. }
\label{fig:selection_field_comparison}
\end{figure*}

\begin{figure*}[t]
\centering
\includegraphics[width=0.98\textwidth]{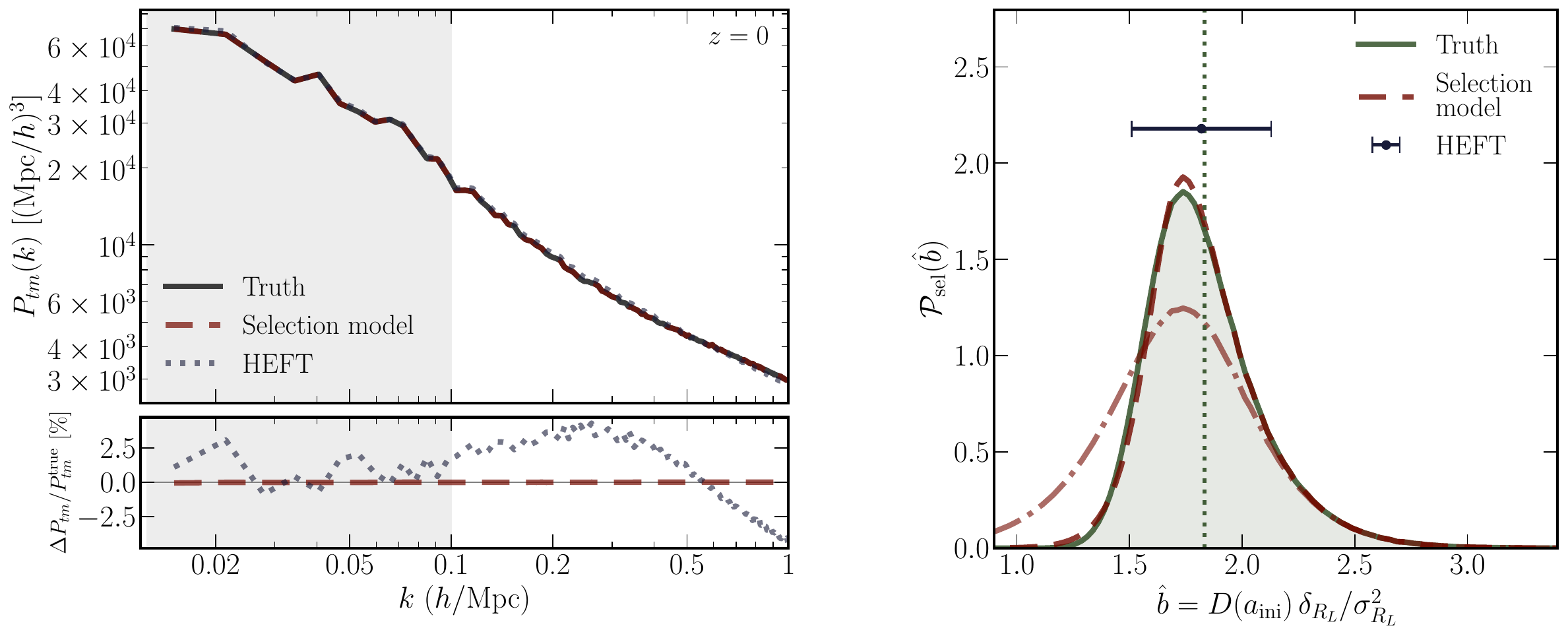}
\caption{\textit{Left panel:} Measurements of the Eulerian ($z=0$) tracer-matter cross $P(k)$ for the truth (Eq.~\ref{eq:fuzzy_truth_selection}), the best-fit probabilistic selection model (Eq.~\ref{eq:logistic_recovery_selection}), and the best-fit HEFT model (Eq.~\ref{eq:heft_lagrangian_tracer_overdensity}), with the fits being performed in Lagrangian space. The HEFT model is a decent fit to the measurement up to $k \simeq 0.1 \, h/{\rm Mpc}$, beyond which there is a visible scale-dependence in the residual. The probabilistic selection model is a good fit down to the smallest scales. \textit{Right panel:} The selected bias distribution from the truth (solid line) and the inferred best-fit (dashed line). The dot-dashed line illustrates the predicted bias distribution when the parameters are not chosen to be a good fit. The solid and dashed lines agree well except at the left tail, while the dot-dashed distribution is quite different. The vertical dotted line represents the mean of the true distribution. The HEFT symbol is centered on the growth-rescaled $b_1$, with a horizontal half-width inferred from $b_2$ by matching the first two bias moments. Its center and width are consistent with those of the true distribution.}
\label{fig:selection_cross_power_comparison}
\end{figure*}

The positivity of the tracer field in Lagrangian space also translates into tracer field positivity in this formalism in Eulerian space. If the displacement field is denoted by $\mathbf \Psi (\mathbf q)$, then the Eulerian tracer overdensity can be written as 
\begin{equation}
\begin{aligned}
1+\delta_{t,\boldsymbol\theta}^{\rm E}(\mathbf x)
&=\int d^3q\,
\frac{p_{\boldsymbol\theta}(\mathbf q)}
{\langle p_{\boldsymbol\theta}\rangle}\\
&\quad\times\delta_{\rm D}^{(3)}
\Big(\mathbf x-\mathbf q-\boldsymbol\Psi(\mathbf q)\Big).
\end{aligned}
\label{eq:positive_advected_selection_field}
\end{equation}
Again the RHS is $\geq 0$ as it is simply a redistribution of the Lagrangian values along the displacements. Note that the displacement field can be calculated from perturbation theory, or represent the full, nonlinear displacement as calculated from an $N$-body simulation~\cite{wu2022}.

We now demonstrate these properties of the bias field and selection formalism, using a slightly modified version of the threshold selection model above. We replace the sharp selection threshold in $\delta_{R_L}$ by a smoother selection
\begin{equation}
p_{\rm true}(\mathbf q)
=\frac12\,\mathrm{erfc}\!\left[
\frac{\delta_{R_L}^c-\delta_{R_L}(\mathbf q)}
{\sqrt{2}\,\tau}\right],
\label{eq:fuzzy_truth_selection}
\end{equation}
where we have chosen $\delta_{R_L}^c$ such that $\langle p_{\rm true}\rangle = 0.01$ and $\tau = 0.2\sigma_{R_L}$. This selection model defines $1 + \delta^L_t(\mathbf q) = p_{\rm true}(\mathbf q)/\langle p_{\rm true} \rangle$. Through advection, using displacements directly from the $N$-body simulation, this also produces an Eulerian overdensity field. Both  Lagrangian and Eulerian density fields are guaranteed to be positive by our construction. We can also compute the Eulerian $P_{tm}(k)$ and the bias distribution of the selected points. These are shown in the two panels of Fig.~\ref{fig:selection_cross_power_comparison}. We will treat this field and its representations in the mentioned figures as the \enquote{true} tracer field measurements for the comparison below.

We now try to recover this \enquote{true} selection by an approximating distribution
\begin{equation}
p_{\boldsymbol\theta}(\mathbf q)
=\left\{1+\exp\!\left[
-\frac{\delta_{R_L}(\mathbf q)-\delta_c}{s}\right]\right\}^{-1},
\label{eq:logistic_recovery_selection}
\end{equation}
with controlling parameters $\boldsymbol\theta=(\delta_c,s)$ and $s>0$. The number of parameters is the same as the actual selection model, but given the different functional forms, the truth and the model will always have location-to-location variation in the probabilities. Given a choice of parameters $\boldsymbol\theta$, Eq.~\ref{eq:logistic_recovery_selection} returns a tracer overdensity field $1+ \delta_{t,\boldsymbol\theta}^L(\mathbf q) = p_{\boldsymbol\theta}(\mathbf q)/\langle p_{\boldsymbol\theta} \rangle$. By minimizing the difference in mean tracer abundance and Lagrangian $P_{tm}(k)$ over the range $2k_F\leq k<0.1\,h\,\mathrm{Mpc}^{-1}$ between the truth and the model we infer the best-fit model parameters. We find that even for the best-fit model, individual pixel values can differ by several percent between the truth and the model.

Using the best fit parameters, we calculate the tracer overdensity field in Lagrangian space, along with the resulting bias distribution. Using displacements directly from the simulation, we calculate the tracer overdensity in Eulerian space, and the Eulerian $P_{tm}(k)$ matter-tracer cross-correlation. These are plotted in Figs.~\ref{fig:selection_field_comparison} and~\ref{fig:selection_cross_power_comparison}. We see that even though the selection model is only an approximation to the underlying true field, the tracer density is always positive. In addition, the Eulerian $P_{tm}(k)$ is an extremely good match to small scales, and the resulting bias distribution only differs from the truth in the left tail - the mean and the width are well-matched. The dot-dashed line in the right panel of Fig.~\ref{fig:selection_cross_power_comparison} is illustrative of the bias distribution obtained for a choice of $\boldsymbol \theta$ that is not the best-fit. The difference is meant to illustrate the fact that inferring the incorrect selection rule will generically lead to an incorrect prediction for the bias distribution, and therefore, clustering of the selected tracers.

To contrast these results against a state-of-the-art benchmark for tracer clustering, we use the hybrid Lagrangian bias or Hybrid Effective Field Theory (HEFT) formalism~\cite[e.g.][]{modi2017,kokron2021,angulo2021,pellejero2022}. This approach uses a perturbative Lagrangian bias expansion, usually truncated at the second order with nonlinear displacements from $N$-body simulations, and has been demonstrated to match tracer clustering to smaller scales than other approaches. The second step of the model, connecting Lagrangian space to Eulerian space via non-perturbative displacements, is, therefore, equivalent to the selection model above. The Lagrangian tracer overdensity can be written as

\begin{equation}
\begin{aligned}
\delta_{t,{\rm HEFT}}^{\rm L}(\mathbf q)
&=b_1^{\rm L}\delta_{R_L}(\mathbf q)
 +b_2^{\rm L}
 \bigl[\delta_{R_L}^2(\mathbf q)-\langle \delta_{R_L}^2\rangle\bigr]\\
&\quad+b_{s^2}^{\rm L}
 \bigl[s_{ij}(\mathbf q)s_{ij}(\mathbf q)
       -\langle s_{ij}s_{ij}\rangle\bigr]\\
&\quad+b_{\nabla^2\delta}^{\rm L}
 \nabla^2\delta_{R_L}(\mathbf q),
\end{aligned}
\label{eq:heft_lagrangian_tracer_overdensity}
\end{equation}
with all operators evaluated at the same
smoothing scale, i.e.,
$s_{ij}=(\partial_i\partial_j\nabla^{-2}-\delta_{ij}/3)\delta_{R_L}$.
The bias coefficients are treated as the free parameters of the model, and we evaluate them by fitting all relevant power spectra $P_{t\mathcal O}$ in the range $2k_F\leq k<0.1\,h\,\mathrm{Mpc}^{-1}$ in Lagrangian space, with $t$ representing the \enquote{true} tracer field from Eq.~\ref{eq:fuzzy_truth_selection}. After the fitting procedure, Eq.~\ref{eq:heft_lagrangian_tracer_overdensity} gives us a field-level representation of the true tracer field within the hybrid Lagrangian bias framework. We can use this representation to compare against both the truth and the field-level representation of the probabilistic selection model.

Fig.~\ref{fig:selection_field_comparison} shows the contrast between the bias field framework and the usual bias expansion. While both are fits to the \enquote{true} tracer overdensity, the underlying objects being fit are completely different, and the two frameworks end up producing very different overdensity fields. In particular, the bias expansion produces a tracer overdensity field that has negative pixel values even in Lagrangian space. When smoothed at $10 \, {\rm Mpc}/h$, $32.8\%$ of all pixels have negative values, i.e. $\delta^L_{t,{\rm HEFT}} \leq -1$. On smaller smoothing scales, an even larger fraction of the pixels have this feature. Advection to Eulerian space increases the fraction of negative pixels to $45.1\%$ of all pixels when smoothed at the same comoving scale. This is represented in the right panels of Fig.~\ref{fig:selection_field_comparison}, where the HEFT histogram has a prominent tail below $-1$. Middle panels present the same information through projected density slabs. Using a projection of $15.6\, {\rm Mpc}/h$, we plot the $\log(1+\delta_t)$ from the HEFT model, analogous to the selection model in the left panels: the top panel is for the Lagrangian field, while the bottom panel is for the Eulerian field. The fraction of negative pixels in this $2$-d projection is noted at the bottom of each panel. Gray pixels represent the negative density points - it is clear that while the fit is optimized to match tracers which typically live in the high-density tail of $\delta_t$, the underdense regions receive negative density value assignments. Deep voids in Eulerian space exacerbate the issue. Note that while we have truncated this expansion at second order, any perturbative bias expansion is likely to run into this problem in different parts of parameter space, especially for highly clustered tracers.

While the tracer field from the hybrid Lagrangian bias approach yields negative tracer density values, the Eulerian $P_{tm}(k)$ from the model is a good fit to the true $P_{tm}(k)$ up to $k \simeq 0.1 \, h/{\rm Mpc}$. Beyond that the residuals are still small, but there is added scale-dependence. This is shown in the left panel of Fig.~\ref{fig:selection_cross_power_comparison}. The probabilistic selection model and bias field formalism, on the other hand, is a good fit in this toy model to all scales that can be measured on the grid. The right panel shows the underlying reason: the selection model gets almost the entire bias distribution that defines the true selection, while the hybrid Lagrangian bias gets the two lowest moments of the distribution correctly. We reiterate that a bad selection model, or parameter values yielding a bad fit to the true underlying selection (dot-dashed line) will typically yield an incorrect prediction for the bias distribution, bias coefficients, and hence, the observed clustering.

We have demonstrated how the bias field formalism, in addition to a selection rule, yields the resulting large-scale clustering of a selection of tracers. While the selection rule presented here was extremely simple to focus on the strengths and features of the new formalism, the general principle remains the same in the presence of more complicated selection or collapse criteria involving multiple initial fields. A proposed selection criterion, written in terms of operators of the initial Lagrangian overdensity, sets the selected bias distribution through the joint statistics of these operators with the bias field. This, in turn, fixes the bias parameters, which are expressed as appropriate moments of the distribution.

If there are additional active operators, we need to generalize Eq.~\ref{eq:selected_bias_statistics}. Specifically, if the orthogonalized operator $\mathcal O_\perp$ is active in selection, then we can write 
\begin{equation}
\left\langle b_{\mathcal O_\perp}\right\rangle_{\rm sel}
=\frac{1}{\sigma_{\mathcal O_\perp}^{2}}
\int d\hat b\,\mathcal P_{\rm sel}(\hat b)\,
\left\langle\mathcal O_\perp\mid\hat b \right\rangle_\mathrm{sel}, 
\label{eq:secondary_selected_bias_integral}
\end{equation}
where $\sigma_{\mathcal O_\perp}^{2}$ is the variance of the operator calculated over the full volume. Therefore, given a selection, the secondary bias coefficients are obtained by evaluating the correctly normalized expectation of the orthogonalized operator over the selected bias distribution. It is easy to see that if we average over the full volume, or if selection depends only on density (independent of $\mathcal O_\perp$), the integral evaluates to $0$.

The bias field formalism, therefore, shifts the focus from large-scale bias coefficients, as currently used, to finding local collapse models in Lagrangian space that statistically capture the collapse of a set of halos or tracers. The local collapse model can be  complicated, and multidimensional, but the general route to clustering predictions from a collapse remains untouched by this degree of complexity. This is yet another way to see the non-perturbative nature of the formalism, and a contrast to the bias expansion approach.

\section{Discussion and outlook}
\label{sec:discussion}

In this paper, we have developed a description of Lagrangian bias as a field defined by the initial Gaussian Random Field of fluctuations, and explored how selections of patches within this field determine the clustering of these tracer patches. We demonstrated how this structure emerges from Gaussian conditioning using both Fourier shell and real space pointwise estimators. We showed how halo bias emerges from the appropriate averaging of this bias field over Lagrangian patches of the proto-halos. We then explored how additional active selection operators, apart from the smoothed density field, can lead to changes in the selected bias distributions, introduce specific scale-dependent Lagrangian bias, and produce assembly-bias-like splitting in the large-scale bias of sub-populations. Finally, we used an explicit density-threshold selection to demonstrate how a selected bias distribution connects to the coefficients in the usual perturbative bias expansion. The example also helped demonstrate other robust features of the framework: positivity of the resulting tracer field even in Eulerian space under nonlinear displacements, and the potential to extend the range of scales to which current bias models can model tracer clustering.

The central result of this paper is that
scale-independent bias is a property of every point in a Gaussian random
field. Importantly, each Fourier shell independently
has the same normalized conditional mean $\langle \delta_k\rangle_{\mathbf q}/\sigma_k^2 = \delta(\mathbf{q})/\sigma^2$. The expression
$b(\mathbf{q}) = \delta(\mathbf{q})/\sigma^2$ captures the result in a local form. This is an exact statement about the Gaussian field, and not the leading term of an expansion, and the test in Sec.~\ref{sec:bias_measurement} explicitly illustrates convergence toward this field at fixed grid resolution. This leads to a fundamentally different perspective on Lagrangian bias: it is a field, whose statistics are laid down by the physics of primordial fluctuations, and tracer clustering or bias arises from the geometric selection of patches where collapse will happen, rather than being coefficients generated by the collapse process itself. 

In the framework, a specified selection determines the Lagrangian bias distribution and tracer field directly. These can be evolved directly under gravity, with no truncation of terms in an expansion. The traditional bias coefficients, on the other hand, attempt to first summarize the distribution by the first few moments. Neglecting higher moments leads to an explicit truncation. A truncation in Lagrangian space implies a truncation in Eulerian space as well. In this sense, the current framework is non-perturbative in nature in terms of the treatment of selection and clustering. Note that the gravitational evolution being perturbative or non-perturbative, i.e. taken from a simulation, is an independent choice.

There are several connections between our work and the existing approaches in the literature. PBS and separate-universe approaches typically approach the bias coefficients from the change in their population as a response to long-wavelength perturbations in different operators~\cite{schmidt2013,baldauf2011,li2014,wagner2015,barreira2017}. Peaks and excursion-set approaches relate clustering to constraints on the initial fields~\cite{bbks1986,desjacques2010,pressschechter1974,bond1991,sheth2001,maggiore2010,musso2012,musso2012bias,paranjape2012,paranjape2013,robertson2009,desjacques2013png}. These approaches obviously match the current formalism for the leading bias coefficients, derived here from the moments of the distribution. Recent approaches have also demonstrated the approximately Gaussian bias distribution empirically, and have related its moments and cumulants to bias coefficients~\cite{stucker2025a,stucker2025b}. We have shown how this selected bias distribution emerges from the underlying bias field and a selection rule. The Web--Halo Model Peak--Background Split (WHM-PBS) likewise predicts a distribution of Eulerian halo bias at fixed mass from the cosmic-web environments hosting the halos~\cite{brieden2026whmpbs}. The connection to the EFT bias expansion~\cite[e.g.][]{mcdonald2006,assassi2014,desjacques2018,Castorina:2016pqq}, is through the particular operators that characterize tracer selection and their associated coefficients. In the current framework, additional active operators act as additional conditioning variables. The joint statistics of these with the bias fields determine the resulting distribution and the coefficients emerge as operator-weighted averages over the selected populations.

However, the field-level description of bias enabled by the current framework has no clearly defined counterpart in the literature. The Lagrangian bias field is defined before any tracer population or selection rule is chosen. It is precisely because of this field structure that specifying a local (in Lagrangian space) selection rule automatically makes a deterministic prediction for the clustering of all such selected locations, through the resulting bias distributions and moments\footnote{It also assigns bias estimates to individual patches, connecting to measurements of halo-to-halo bias variation~\cite{paranjape2018}.}. The spatial variations in the different relevant fields allow a statistically correct selection prescription to predict the large-scale clustering with no independently fitted bias coefficients. The parameterization now lives in the collapse model. This establishes local collapse modelling as an alternative route to predicting the tracer field and its clustering, with the bias distribution and coefficients emerging as summaries of the selected population. Other aspects that are known to affect clustering, such as stochasticity, also need to be expressed in the language of the current framework.

Additionally, the probabilistic selection model framework in Sec.~\ref{sec:tracer_field_postivity} yields positive tracer field densities in Lagrangian and Eulerian space. This positivity, crucially, is preserved even for an approximate selection model as the selection parameters themselves are fitted. The accuracy of its clustering predictions depends on how well it captures the target true population. While this was demonstrated for a particularly simple selection rule, the general ideas will hold for more complicated, multivariate selection rules. The challenge will be to find optimal ways to parameterize these potentially complicated rules as a function of halo mass or other tracer properties. The comparison to the hybrid Lagrangian bias approaches ~\cite[e.g.][]{modi2017, kokron2021, angulo2021, pellejero2022} shows that these methods can match Eulerian tracer clustering at the power spectrum level while having a large fraction of negative pixels in their field-level representation. The guaranteed tracer density positivity of the current framework is likely to be crucial for modeling higher-order statistics that are sensitive to the full morphology of the Eulerian cosmic web. Statistics such as the $k$-nearest-neighbour cumulative distribution functions ($k$NN-CDFs)~\cite{banerjeeabel2021knn,banerjeeabel2021cross,banerjeeabel2023field}, that are sensitive to the geometry of tracer distributions~\cite{gangopadhyay2025geometry}, and have existing HEFT implementations~\cite{Banerjee:2021cmi,chand2026knn} will be ideal rigorous testbeds.

Several additional next steps arise naturally from this framework. The first and most important will be to develop realistic statistical collapse prescriptions for different halo and tracer populations. We will have to identify the relevant initial operators and selection rules for populations selected by mass or other small-scale properties. Building such flexible parameterizations will enable a test of this framework for cosmological inference. There are potentially beneficial features: as made explicit in Eq.~\ref{eq:probabilistic_selected_bias_distribution}, the selection parameters control the entire selected bias distribution, and therefore changes affect all moments or bias coefficients simultaneously. Degeneracy directions between parameters of the selection model and cosmological parameters are likely to be different from those in the usual bias expansion. A second direction is to quantify the spatial mixing of the Lagrangian bias labels during nonlinear evolution, mapping out the process as a function of resolution. It will be particularly interesting to study the mixing of the Lagrangian bias labels inside virialized structures. A third scientifically interesting direction will be to extend the bias field formalism to cosmologies with multiple relevant matter fields, such as cosmologies with massive neutrinos or additional lighter species. These cosmologies are known to exhibit additional scale-dependence in their bias, and this behavior can be exploited to place more stringent constraints on the properties of the particle species~\cite{loverde2014,castorina2014,banerjee2016,villaescusa2018,banerjee2020,Bayer:2021kwg}. This is already being explored in the traditional bias expansion formalism~\cite{celikschmidt2026mixed}.

\begin{acknowledgments}
 We thank Susmita Adhikari and Aseem Paranjape for helpful discussions. AB’s work was partially supported by grants SRG/2023/000378 and ANRF/ARGM/2025/000301/TS from the Anusandhan National Research Foundation (ANRF) India.

\end{acknowledgments}

\appendix

\section{Bias of non-halo Lagrangian patches}
\label{app:nonhalo_bias}

The patch formalism of Sec.~\ref{sec:tracer_populations} also applies to
populations of locations that do not belong in collapsed regions or halos in the Eulerian field. We illustrate this with
spherical patches of radius $R_L=4.5\,{\rm Mpc}/h$ on the initial
Lagrangian lattice of the same simulation. The selection field
$\delta_{R_L}$ uses the sharp Fourier cutoff $k<1/R_L$ of
Eq.~\eqref{eq:smoothed_estimator}, and the  variance $\sigma_{R_L}^2=\langle\delta_{R_L}^2\rangle_V$ is evaluated
over the full initial volume, before selection.
We randomly choose $2\times10^5$ patch centers in each of the
$79.5$--$80.5$ and $89.5$--$90.5$ density percentile bands, and include all particles lying inside these patches.

Each patch contains an equal number of particles, so for a population $\mathcal S$ of $N_{\mathcal S}$ patches, we use
\begin{equation}
\hat b^L_{\mathcal S}
=\frac{1}{N_{\mathcal S}}\sum_{p\in\mathcal S}
 \frac{1}{N_p}\sum_{i\in p}\hat b_{\mathcal K}(\mathbf q_i),
\label{eq:nonhalo_population_estimator}
\end{equation}
where $\hat b_{\mathcal K}$ is Eq.~\eqref{eq:optimal_estimator}, using
shells with mean wavenumbers $0.02\leq k<0.05\,h/{\rm Mpc}$. The summation over $i$ represents particle summation in a patch, while the summation over $p$ represents the summation over patches.
The advected prediction in Eulerian space is then given by
\begin{equation}
\hat b^{\rm pred}_{E,\mathcal S}(a)
=1+\frac{D(a_{\rm ini})}{D(a)}\hat b^L_{\mathcal S}.
\label{eq:nonhalo_eulerian_prediction}
\end{equation}

\begin{figure}[t]
  \includegraphics[width=0.98\columnwidth]{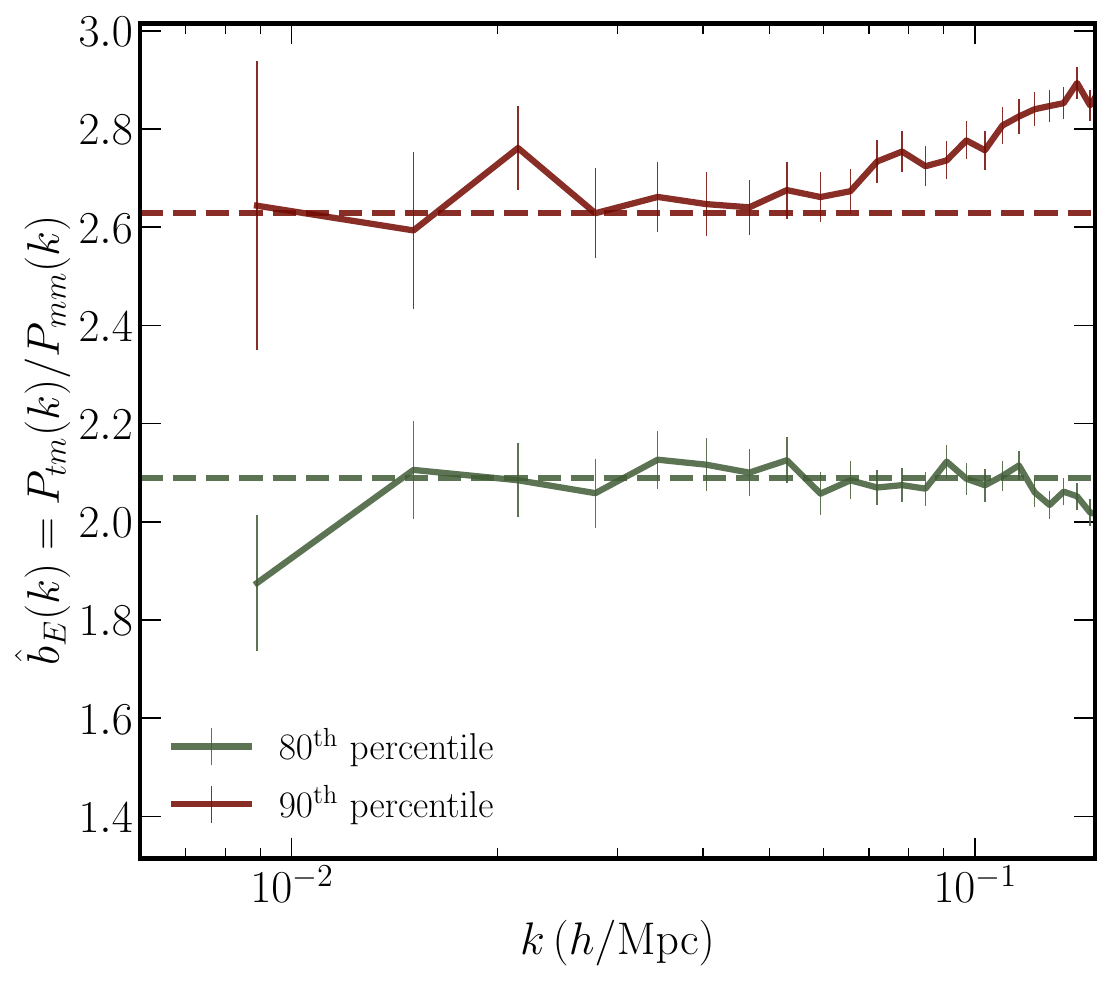}
  \caption{Eulerian bias measurements for two
  density-selected patch populations, with no reference to halo membership, described in
  Appendix~\ref{app:nonhalo_bias}. Solid curves show the measurement of $P_{tm}/P_{mm}$. Dashed lines show the bias field predictions from
  Eq.~\eqref{eq:nonhalo_eulerian_prediction}. }
  \label{fig:particle_eulerian_split}
\end{figure}

We compare the bias prediction against the actual measurement of the clustering of these particles advected to their Eulerian locations in Fig.~\ref{fig:particle_eulerian_split}. Similar to the demonstration with halos in the main text, the bias field prediction is a good match to the clustering on large scales at $z=0$. This establishes the validity of the formalism over selections that are not necessarily collapse prescriptions.

\section{Secondary bias in non-halo patches}
\label{app:nonhalo_secondary}

We next impose a selection depending on both $\delta_{R_L}$ and
$\nabla^2 \delta_{R_L}$ to illustrate the connection between
selection and clustering discussed in Sec.~\ref{sec:halo_assembly}, but for non-halo patches.
Using the same $R_L=4.5\,{\rm Mpc}/h$,
we randomly select $2\times10^5$ centers in the $75$th--$80$th
percentile of $\delta_{R_L}$ and $10$th--$25$th percentile of
$\nabla^2\delta_{R_L}$. We then average density and curvature
over the particles in each patch. Sorting by patch-averaged density, we create $40$ bins of $5000$ objects each. Within each bin,
we divide at the median patch-averaged curvature and pool the two populations across bins, obtaining $10^5$ patches in each final sub-population.
The two mean patch densities agree at the level of $\sim 10^{-6}$, as intended in our design.

We measure the Eulerian bias of the full population and the two sub-populations at $z=0$ and plot them with the solid lines in Fig.~\ref{fig:particle_split}. The bias predictions are calculated for each population as outlined in Appendix~\ref{app:nonhalo_bias}, where we have already restricted ourselves to $k$-shells where $k\ll 1/R_L$. The bias field predictions are plotted for each population with dashed lines. On large scales the measurements and predictions agree, establishing that the arguments about secondary halo bias in the main text generalize to other selections in the Lagrangian field.

\begin{figure}[t]
  \includegraphics[width=0.98\columnwidth]{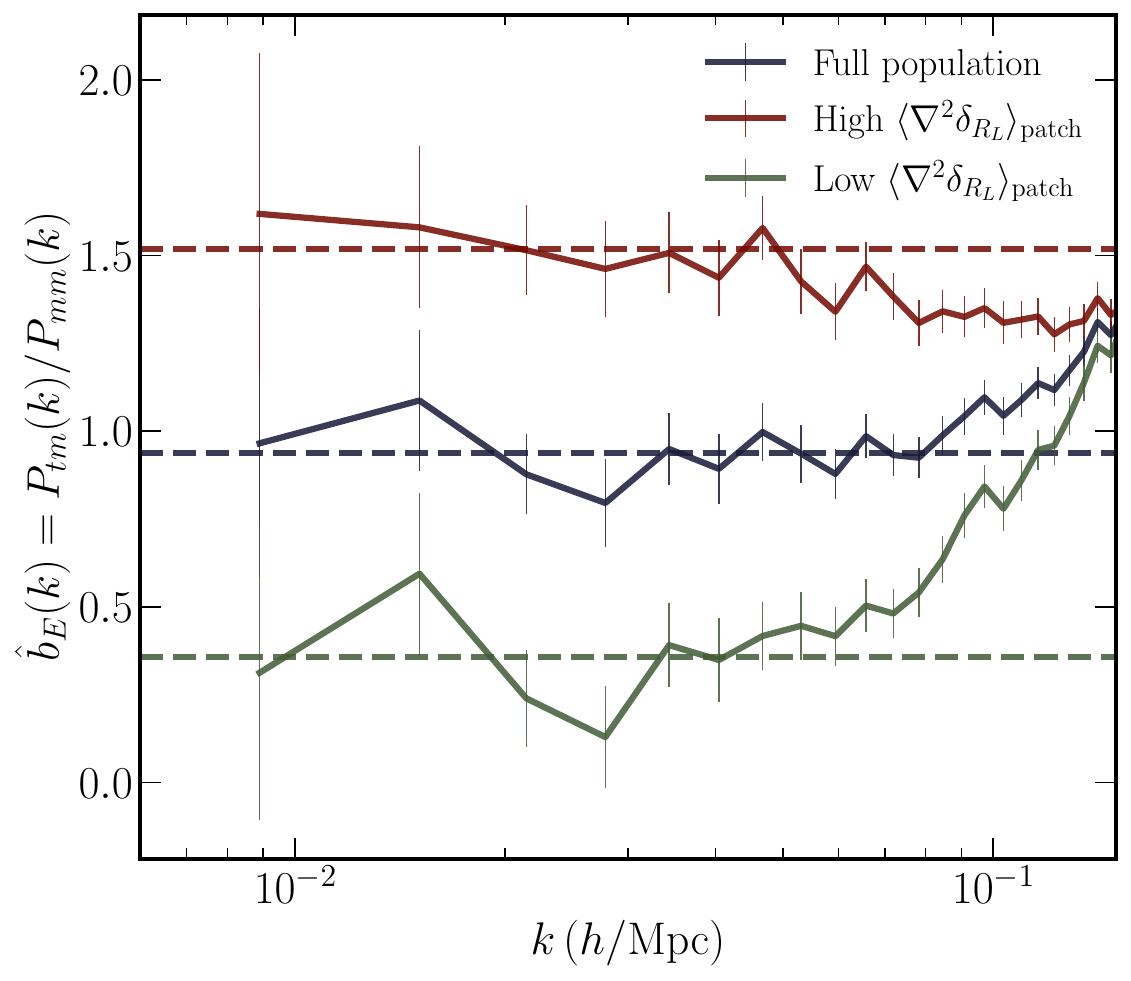}
  \caption{Eulerian bias measurement for the joint
  density and curvature selection given in  Appendix~\ref{app:nonhalo_secondary}.
  Solid lines show the measurement of $P_{tm}/P_{mm}$ at $z=0$ while dashed lines show
  Eq.~\eqref{eq:nonhalo_eulerian_prediction} evaluated separately for
  each population. The split in the large scale bias of the two sub-populations is seen as prominently as for halo patches.}
  \label{fig:particle_split}
\end{figure}

\bibliography{refs.bib}

\end{document}